\documentclass[agupp]{aguplus}

\renewcommand{\arraystretch}{1.2}
\usepackage{epsfig}
\input{extrasym.sty}
\usepackage{bencmmib57}
\lefthead{Baumert}
\righthead{Turbulence by shear \& waves ..}
\sectionnumbers
\printfigures
\doublecaption{35pc}
\epsfverbosetrue 

\authoraddr{H. Z. Baumert, IAMARIS, Bei den M\"uhren 69A,\\
D-20457 Hamburg, Germany (baumert@iamaris.org)}

\slugcomment{Report on the SHEWAMIX project, Grant N62909-10-1-7050, ONR-Global. }

\printfigures
\renewcommand{\fboxrule}{0pt}

\begin{document}
\bibliographystyle{ametsoc}

\title{Turbulent mixing driven by mean-flow shear and internal gravity waves in oceans and atmospheres}

\author{Helmut Z. Baumert}
\affil{Institute for Applied Marine and Limnic Studies, Hamburg, Germany}

\begin{abstract}
This study starts with balances deduced by Baumert and Peters (2004, 2005) 
from results of stratified-shear experiments made in channels and wind tunnels 
by Itsweire (1984) and Rohr and Van Atta (1987),
and of free-decay experiments in a resting stratified tank by Dickey and Mellor (1980). 
Using a modification of Canuto's (2002) ideas on turbulence and waves, these 
balances are merged with an (internal) gravity-wave energy balance 
presented for the open ocean by Gregg (1989), without mean-flow shear.
The latter was augmented by a linear (viscous) friction term. 
Gregg's wave-energy source is interpreted on its long-wave spectral end as internal 
tides, topography, large-scale wind, and atmospheric low-pressure actions. In addition, 
internal eigen waves, generated by mean-flow shear, and the aging of the wave field from 
a virginal (linear) into a saturated state are taken into account. Wave packets and 
turbulence  are treated as particles (vortices, packets) by ensemble kinetics so that
the loss terms in all three balances have quadratic form. Following a proposal by Peters (2008), the
mixing efficiency of purely wave-generated turbulence is treated as a universal constant,
as well as the turbulent Prandtl number under neutral conditions. 
It is shown that: (i) in the wind tunnel, eigen waves are switched off, 
(ii)  due to remotely generated long waves or other non-local energy sources, coexistence equilibria of turbulence and 
waves are stable even at Richardson numbers as high as $10^3$; 
(iii) the three-equation system is compatible with geophysically shielded settings like certain stratified laboratory flows.
The agreement with a huge body of observations surprises. Gregg's (1989) wave-model component 
and the a.m. universal constants taken apart, the equations contain only one additional dimensionless
parameter for the eigen-wave closure, estimated as $Y\approx 1.35.$
\end{abstract}


\section{Introduction\label{intro}}
\subsection{General\label{introgeneral}}

\noindent
In geophysical flows, turbulence is ubiquitous. 
Today turbulent engineering flows may already be simulated 
using DNS, i.e.\ fully resolving scales down to the smallest ones, provided 
they are large compared with molecular cluster scales. However, for ocean and 
atmospheric flows this will remain impossible even in the foreseeable 
future. Here particularly the dominating stably stratified flows, 
i.e.\ all forms of coexistence of turbulence and 
internal waves, represent a specific observational and theoretical challenge.
Without a better understanding of the fundamental physics of these 
processes all our regional to global models of weather and climate remain incomplete. 

Oceanic, atmospheric and stellar turbulence has challenged a number of prominent scientists and research 
groups since long, in the recent decade namely Woods (2002), Galperin et al. (2007), 
Canuto et al. (2008), and Zilitinkevich et al. (2008), also Kantha and Carniel (2009).
They all emphasized a major  contradiction between observational experience and existing theories: 
\begin{itemize}
  \item[(i)] For controlled stratified shear flows in the laboratory exists undoubtly a critical 
  gradient Richardson number of $R_g^b \le {\cal O}(\onequarter)$ above which turbulence dies out 
(Itsweire, 1984; Itsweire et al., 1986; Rohr and Van Atta, 1987; Rohr et al., 1987; Rohr et al., 1988a, b; Van Atta, 1999). 
Also in the field the qualitative and strongly (but not totally) limiting role of $R_g^b$ 
is without doubt (Peters et al.,  1988) and became specifically visible with the advent of Lagrangian floaters 
in the Lagrangian time spectra of turbulence and wave-like fluctuations where $\Omega=N$ marks a 
sharp divide: to the left a flat wave spectrum, to the right a Kolmogorov time spectrum 
(D'Asaro and Lien, 2000a, b). 
Also the existing theories (Richardson, 1923; Miles, 1961; Howard, 1961; 
Hazel, 1972; Thorpe, 1973, Abarbanel et al.,  1984; Baumert and Peters, 2004) 
point all into the same direction.

\item[(ii)] Geophysical flows are reported to exhibit not always, but more than often significant stable turbulence levels 
	and mixing capabilities  at $R_g \gg \onequarter$ (Peters et al., 1988; Canuto, 2002; Poulos et al., 2002;
	Nakamura and Mahrt, 2005; Grachev et al., 2005, 2006). Also Peters and Baumert (2007) report 
	problems in validating a $K$-$\varepsilon$ turbulence closure against comprehensive estuarine	
	microstructure measurements in the Hudson river. 
	For tidal phases with weak shear the observed turbulence levels
	exceeded the model values by 2 to 3 orders of magnitude. The dynamic time lag of the turbulent state variables 
	in the M$_2$ tide (about one hour, see Baumert and Radach, 1992) can be excluded as a source of the deviations
	because the model naturally contains this effect. 
\end{itemize}
The text below makes an attempt to resolve the sketched contradiction using ideas of Woods (2002 and literature
cited therein) and Canuto (2002) regarding the role of waves and billow turbulence. We identify internal gravity waves 
(so far neglected in most other studies) as the potentially responsible phenomenon.

For a better understanding of the language used in this text we first introduce   notational conventions 
and  then continue with an overview of major physical processes within the world of 
stratified shear turbulence and internal gravity waves. 

Below, \textit{we} always means  the author and the dear reader in a dialogue.


\subsection{Setup and notation \label{introsetupnotation}}

\noindent
For  simplicity  we focus on a  simple non-trivial situation, 
a spatially one-dimensional ``channel'' flow with velocity component $U$ in horizontal 
($x$)  direction, with variation of $U$ along the vertical, $z$ (pointing upwards), 
and with Eulerian density and icopycnal coordinate fluctuations. 

The  decomposition of our flow field into mean and fluctuations reads as follows:
\begin{eqnarray}\label{perturb1}\label{U}
	U(z, t) &=& \langle U\rangle + \tilde u(z,t) + u'(z,t)\,,\\
	\label{V}
	W(z, t) &=& \langle W\rangle + \tilde w(z,t)+w'(z,t)\,.
\end{eqnarray}
Variables with tilde are small-scale short-wave components\footnote{actually wave packets}, present only under stratified conditions.
Primed variables denote turbulent fluctuations. Both fluctuating components vanish in the mean. 

The \textit{turbulent} kinetic energy, TKE or $\cal K$, is defined as follows,
\begin{equation}\label{tke}
	{\cal K} = \frac{1}{2} \langle u'^2+v'^2+w'^2\rangle.
\end{equation}
A wave's total energy, $\cal E$, is the sum of potential, $\langle N^2\, \tilde \zeta^2\rangle/2$,  
and kinetic energy, $\langle\tilde u^2+\tilde w^2\rangle/2$  (Gill, 1982):
\begin{equation}\label{wte}
	{\cal E} =  \frac{1}{2} \langle \tilde u^2+\tilde w^2+N^2\, \tilde \zeta^2\rangle.
\end{equation}
Here $\tilde u$ points into the direction of wave propagation so that
in a plane wave $\tilde v =0$. Average $\langle . \rangle $ is taken at least over one wave period
which ``by definition''  is longer than the characteristic turbulent time scale. In the following the 
dimensionless gradient Richardson number, $R_g$, plays a central role. 
In a stra\-ti\-fied shear flow as above, shear is given by 
\begin{equation} \label{S2}
	\langle S\rangle = \frac{\partial \langle U\rangle}{\partial  z}\,,
\end{equation}
while stratification is characterized by the Brunt-V\"ai\-s\"a\-l\"a frequency squared, 
\begin{equation} \label{N2}
	\langle N^2\rangle = -\frac{g}{\langle \rho\rangle} \,\frac{d\langle\rho\rangle}{d z}.
\end{equation}
$\langle\rho\rangle$ is the background density field. 

The gradient Richardson number, $R_g$, characterizes the dimensionless ratio of the two
aspects, shear and stratification:
\begin{equation}\label{rg}
R_g=\langle N^2\rangle /\langle S\rangle^2\,.
\end{equation}
Below the averaging operators are mostly omitted for brevity of notation.
But fluctuations are consequently labeled either by tilde (wave-like) or prime (turbulent).

Based on $R_g$, under \textit{controlled laboratory} 
conditions  where the linear  eigen waves leave 
the experimental site before quadratic saturation 
and a feed back into the TKE pool can happen
(wind tunnel of Van Atta),
the following hydrodynamic regimes for horizontally  
homogeneous flows are found (Baumert and Peters, 2004, 2005):
\begin{itemize}
\item[(a)] $R_g \le R_g^a\equiv 0$: unstable and neutral stratification,\\
	convective turbulence, no internal waves at all.
\item[(b)] $0 \equiv R_g^a < R_g < R_g^b \equiv  \onequarter$: stable stratification, 
	shear-dominated growing turbulence, coexistence of turbulence and internal waves.
\item[(c)] $ \onequarter \equiv R_g^b < R_g < R_g^c \equiv  \onehalf$: stable stratification,\\
	wave-dominated decaying turbulence, coexistence of turbulence and internal waves.
\item[(d)] $\onehalf \equiv R_g^c < R_g $: stable stratification, no coexistence of turbulence 
	and waves, \textit{waves-only} regime.
\end{itemize}
Under those conditions the turbulent Prandtl number $\sigma$ is (Baumert and Peters, 2004, 2005)
\begin{equation} \label{prandtl1}
	\sigma = \frac{\mu}{\nu} = \frac{\sigma_0 }{1- (N/\Omega)^2}
\end{equation}
The above values for the critical numbers $R_g^a, R_g^b$ and $R_g^c$
hold for the asymptotic case $Re \rightarrow\infty$. 

The situation is different if we leave the laboratory wind tunnel and consider the
open ocean, stratified rivers or the stably stratified atmosphere where turbulence is 
not only locally generated through local mean-flow shear but also through 
the action of space-filling (non-local) spectra of internal gravity waves (IGWs).
These are generated e.g.\ by tidal forces, possibly at remote places, arriving 
at our point of interest along various pathways.


\section{Major physical interactions in stably stratified oceans and atmospheres\label{physics}}

\begin{figure}[h] 
\begin{center} 
\includegraphics[width=4cm,height=6cm,keepaspectratio]{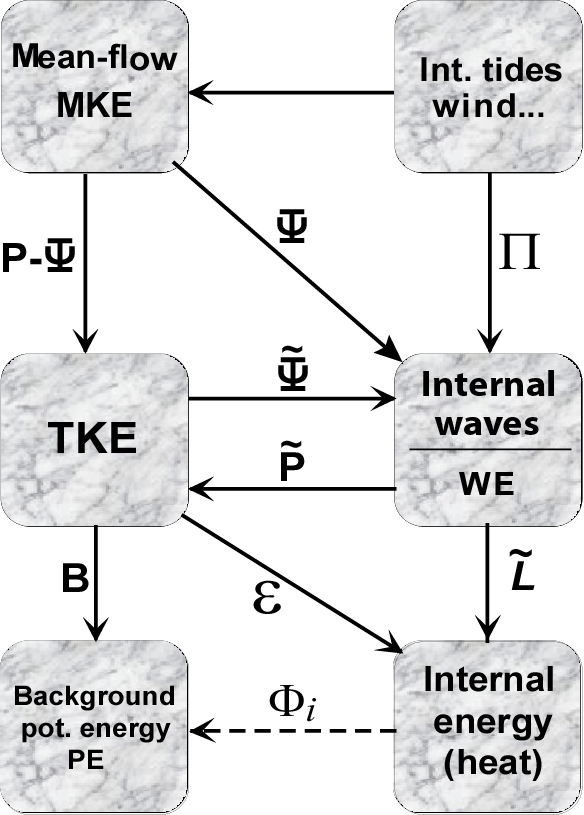}
\end{center}
\caption{Interaction of major forms of fluid-mechanical energy in 
stratified oceanic, atmospheric or stellar shear flows. $\Psi$ is the flux of eigen-wave energy mentioned
in the text. $\Pi$ is the energy flow which corresponds to remotely generated waves. The wind tunnel 
allows to cut off the flux $\tilde P$ from the the wave-energy to the TKE pool because the wave spectrum
cannot reach a saturated state. In the long-term equilibrium holds $\tilde P=(1-f) (\Psi+\Pi)$ where $f$ is the
fraction of direct wave dissipation and tends to zero if the wave spectrum approaches saturation.
$f$ may become relevant for short wave ages. }
\label{f:energy}
\end{figure}

\begin{figure}[h] 
\begin{center} 
\includegraphics[width=5.5cm,height=5cm,keepaspectratio]{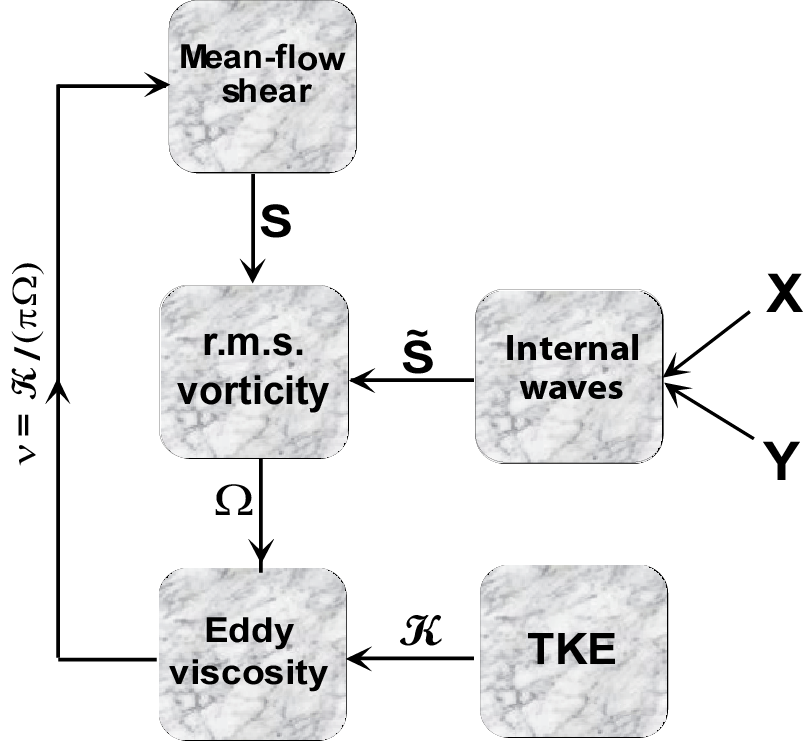}
\end{center}
\caption{Interaction of major vorticity-controlling motion components in 
stratified oceanic and atmospheric shear flows. The r.m.s. vorticity $\Omega$ is governed by the mean-flow shear, $S$,
and a wave-induced pseudo shear, $ \tilde S$. The latter is controlled by eigen waves ($Y$) and remotely generated long waves ($X$).
For details see text.}
\label{f:vorticity}
\end{figure}

\noindent
Fig.\ \ref{f:energy} schematically presents the major interactions between the energies of 
mean\footnote{Mean-flow kinetic energy, $\onehalf\,\langle U\rangle^2$, MKE; Internal tides} and fluctuating 
motion components\footnote{TKE, $\cal K$; r.m.s.\ vorticity, $\Omega$; WKE.}.
Fig.\ \ref{f:vorticity}  does the same for r.m.s. vorticity and turbulent viscosity, $\nu$. 
The latter connects $\cal K$ and $\Omega$ with the mean flow through 
the so-called Kolmogorov-Prandtl relation in the following form (Baumert and Peters, 2004, Baumert, 2012):
\begin{equation} \label{nu}
	\nu = {{\cal K}}/{\pi\,\Omega} \,.
\end{equation}
The TKE dissipation rate, $\varepsilon$, transforming TKE into heat, is defined as (Baumert and Peters, 2004, Baumert, 2012)
\begin{equation} \label{eps}
	\varepsilon = {{\cal K}\,\Omega}/{\pi} =\nu \,\Omega^2\,.
\end{equation}
The buoyancy flux, $B$, transforming TKE into background potential energy, $PE_b$,
may be expressed using the eddy diffusivity, $\mu$, as follows\footnote{The buoyancy flux (\ref{BB}) refers to 
purely shear-generated turbulence (for details see Subsection \ref{shearless1} and Baumert and Peters, 2004, 2005).},
\begin{equation}\label{BB}
	B=- \frac{g}{\langle\, \rho\rangle}\langle w'\,\rho'\rangle =\mu\,N^2\,,
\end{equation}
where $\mu$ is  related with the eddy viscosity $\nu$ through $\sigma$,
the turbulent Prandtl number function:
\begin{equation}
\label{mu}
	\mu =\nu/\sigma\, \quad\quad \mbox{or}  \quad\quad \sigma=\nu/\mu \,.
\end{equation}
Finally, the molecular heat flux, $\Phi_i$, transforms internal energy into background potential energy, $PE_b$.

There is further the rate $\Pi$, which transforms the energy of internal tides and wind-generated large-scale internal 
motions over a long chain of various friction-poor wave-wave interactions into short-wave internal-wave energy, 
and there is the rate $\tilde P$, which transforms the energy of internal-waves spectra into TKE, mainly by breaking 
(a shot noise process), and to a low degree by wave shear.

The mean-flow shear, $S$, controls TKE production through $P-\Psi$ and  internal-wave generation through $\Psi$. 
In the (conventional) neutrally stratified case, $N^2 =0$, the total loss term in the mean-flow kinetic-energy balance 
(MKE) is $-P$, which is at the same time the only source of small-scale energy:  
\begin{equation} \label{P}
	P= \nu\, S^2\,.
\end{equation}
But for $N^2>0$ small-scale energy means the \textit{sum} of turbulence \textit{and} waves, $P=(P-\Psi)+\Psi.$

The shear governs not only TKE and IGW production but also the generation of r.m.s.\ vorticity. But vorticity is also influenced by short 
internal-gravity waves through the term $\tilde S$: in the shearless IGW-dissipation case (e.g.\ Gregg, 1989) we have also vorticity generation.

\noindent
When we use  the word \textit{energy} in the present context, we mostly mean 
for brevity TKE ($\cal K$), i.e.\ turbulent kinetic energy. If we talk here about \textit{vorticity} we similarly mean for 
brevity the r.m.s. turbulent vorticity. For its detailed mechanical interpretation we refer to Baumert (2012).

The precise meaning of $\Omega$ is actually the module of the \textit{vorticity frequency}.
The module of the vorticity itself, $\omega$,  is  $\omega = 2\,\pi\,\Omega$. 
The enstrophy is ${\omega^2}/{2}$.


\section{Balances for small-scale motions\label{balance}}

\noindent
Before we start to write down transport equations for primary variables we first discuss their nature and their balances. 
For this aim we look again at Fig.\ \ref{f:energy} and there namely on the two boxes in the middle row with the names 
\textit{TKE} and \textit{Internal waves}. 

The left box, \textit{TKE}, is fed by two components: by mean-flow shear in the form of $P-\Psi$, and by 
internal-wave breaking, $\tilde P$, see (\ref{tildeP}) below. But it generates heat by $\varepsilon$ via (\ref{eps}), exports buoyancy $B$ via 
(\ref{BB})\footnote{generating thus background potential energy, $PE_b$.}, and generates by $\tilde \Psi$ in the course of aging and 
subsequent collapse short internal waves, when its time scale approaches the internal-wave period
(Baumert and Peters, 2005). 

Turbulence collapsing into waves is an event mostly bound to certain special conditions 
(e.g.\ Dickey and Mellor, 1980;  D'Asaro and Lien, 2000a, b). Under smooth and almost-equilibrium conditions 
the TKE may collapse and generate waves which then after aging saturate and feed their energy back
into the TKE pool, as part of $\tilde P$. In the following $\tilde \Psi$ is therefore mostly neglected.

The right box, \textit{Internal waves}, is fed by three components: by large-scale  wave sources, $\Pi$, 
by the rate of eigen-wave generation, $\Psi$, and by the collapse rate discussed already above and 
neglected further below. This box further exhibits two relevant losses: the linear wave friction, $\tilde L$, 
\begin{equation} \label{tildeL}
	\tilde L = c_1\, {\cal E} \,,
\end{equation}
and the \textit{quadratic} wave friction, $\tilde P$, 
\begin{equation} \label{tildeP}
	\tilde P = c_2\, {\cal E}^2 \,.
\end{equation}
The latter is mainly caused by wave breaking and dominates the wave-energy balance, but not completely. 
The loss term\footnote{see also the wave-energy balance (\ref{w-1}) below} (\ref{tildeL}) and (\ref{tildeP})
follows arguments developed by Gregg (1989) which we augment as follows. 

From a purely theoretical point of view the phenomenon of wave breaking tells us nothing about its kinetics. 
It is the processes \textit{before} breaking occurs which form the bottleneck. If we accept the idea that not only 
vortices (actually: vortex-dipole filaments, see Baumert, 2012) are \textit{particles} which move in space until 
collision, then the interpretation of wave-energy dynamics in terms of particle dynamics 
does not come as a surprise, in particular in view of the billow turbulence discussed by Woods (2002). 

Due to the general particle-wave dualism of field theories, which is a well established concept in classical continuum 
mechanics\footnote{It became most famous in quantum mechanics and has been somewhat monopolized there.}, 
also wave packets  can be treated as particles moving with their group velocity until collision. After collision they either
move ahead, or change their paths, or, with low probability, their energy is dissipated by breaking in dissipative patches (billows). 
In contrast to vortex kinetics, here the probabilities are not symmetric. In both cases (vortex dipoles and wave packets) the 
collision events are highly intermittent. 

Besides its phenomenological basis, the quadratic term in (\ref{tildeP}) 
might therefore have deeper roots or at least analogies in the kinetics of reactive particle ``gases'' or 
molecular reactions in fluids where particle-collision probabilities follow the product of their spatial densities. 
For collisions between particles of same kind, the quadratic collision term is thus a logical consequence. 

Fig.\ \ref{f:vorticity} shows the major feedback loop between the small-scale motions and the mean flow
which eventually smoothes the flow through the turbulent viscosity. The central left box,
\textit{r.m.s. vorticity},  is fed by two components: by the mean-flow shear, $S$, and by the internal-wave 
field through the pseudo shear, $\tilde S$. The vorticity $\Omega$ itself controls together with 
the TKE $\cal K$ the turbulent viscosity via (\ref{nu}) which eventually smoothes the flow.

The spectral signatures of shear-generated fluctuations and wave-wave interactions differ qualitatively. 
While shear influences vorticity directly by  prescribing a time scale $\propto S^{-1}$, the long-wave sources ($\Pi$) of IGW 
spectra do it more indirectly via a longer chain of wave-wave interactions cascading down to critical frequencies around $N$. 
This implies that each of the two mechanisms needs ``his'' closure. 

For conditions of homogeneous\footnote{in horizontal and vertical direction} shear, stratification and wave fields
the above three balances can be so far formulated as follows (compare with Figs.\ \ref{f:energy} and \ref{f:vorticity} ): 
\begin{eqnarray}
\label{o-1}
	\frac{d \Omega}{d t}&=&  \frac{1}{\pi}\left(\frac{S^2}{2} - \Omega^2+ \tilde S^2\right),\\
\label{k-1}
	\frac{d {\cal K}}{d t}&=& (P -\Psi) -B - \varepsilon + \tilde P ,  \\
\label{w-1}
	\frac{d \cal E}{dt} &=& (\Pi  +\Psi)-  \tilde L - \tilde P.
\end{eqnarray}
With the exception of $\Pi$, all variables in (\ref{o-1}, \ref{k-1}, \ref{w-1}) have purely local character. 

In (\ref{k-1}) we used the relation
\begin{equation} \label{BW}
	B+\Psi = \nu\, N^2/\sigma_0 \,,
\end{equation}
derived and discussed in greater detail  by Baumert and Peters (2004, 2005). 
Here $\sigma_0=\onehalf$ is the  value of the turbulent 
Prandtl number function $\sigma$ for the case of neutral stratification ($R_g=0$ or $\Omega\rightarrow\infty$).
With (\ref{prandtl1}) we have
\begin{equation} \label{WW}
	\Psi= \frac{\varepsilon}{\sigma_0} \,\left( \frac{N}{\Omega}\right)^4.
\end{equation}
Note that the system (\ref{o-1} -- \ref{w-1}) is not one of the common three-equation models used 
in traditional turbulence modeling. The focus of this modeling branch is directed on higher and higher
orders of the closure equations, e.g.\ equations for second and third moments and so forth, all derived
from the Navier-Stokes equation.  

In later sections we will see that the system 
(\ref{o-1} -- \ref{w-1}) is dynamically stiff. Under spatially homogeneous conditions 
well-defined steady-state solutions exist, but perturbations of this state
are connected with very different characteristic relaxation times of the
systems components. In particular, $\Omega$ is the component relaxing fastest into what
we call structural equilibrium. The TKE ($\cal K$) relaxes significantly slower into
a new  state while the wave energy, $\cal E$, relaxes extremely slowly. The wave-energy
pool needs a longer spin-up time because the flux of shear-generated wave energy, $\Psi$, 
consists of random linear wave packets which simply need time ``to meet and break'', of course by chance.

\begin{figure}[h] 
\begin{center} 
\includegraphics[width=6cm,height=4.5cm,keepaspectratio]{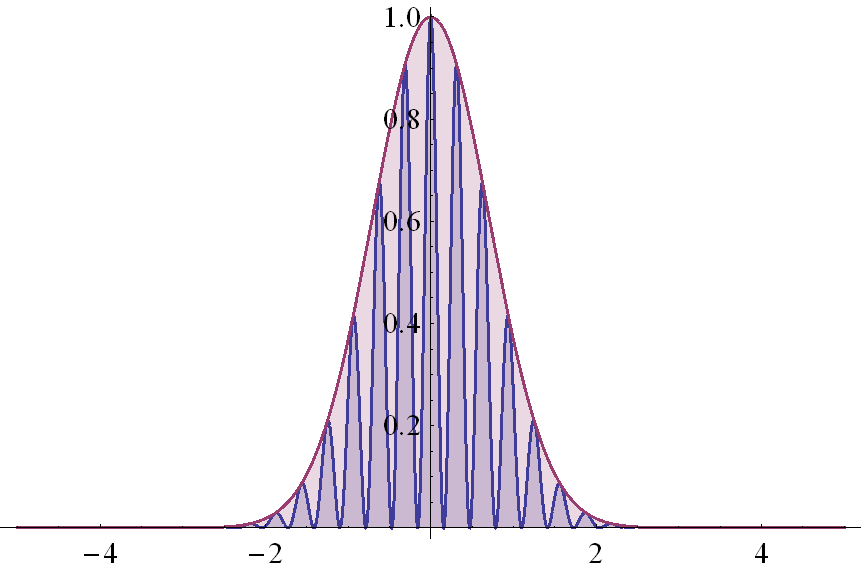}
\end{center}
\caption{Linear wave packet, hypothetically generated by shear. 
Breaking occurs by chance superpositions and results in dissipative patches or billows.}
\label{f:packets}
\end{figure}

For later use we introduce here the ``viscous fraction'' $f$ of the total energy loss of the wave pool
towards the heat pool, 
\begin{equation}\label{f1}
	f=\frac{\tilde L}{\tilde L +\tilde P}
\end{equation}
with the obvious property
\begin{equation}
	f+\frac{\tilde P}{\tilde L +\tilde P}=1.
\end{equation}
The fraction $1-f$ describes a  ``wave age'', i.e.\ the relative wave-energy loss into the TKE pool.
Clearly, 
\begin{equation}
	0\le f <1.
\end{equation}

\begin{figure}[h] 
\begin{center} 
\includegraphics[width=6cm,height=4cm,keepaspectratio]{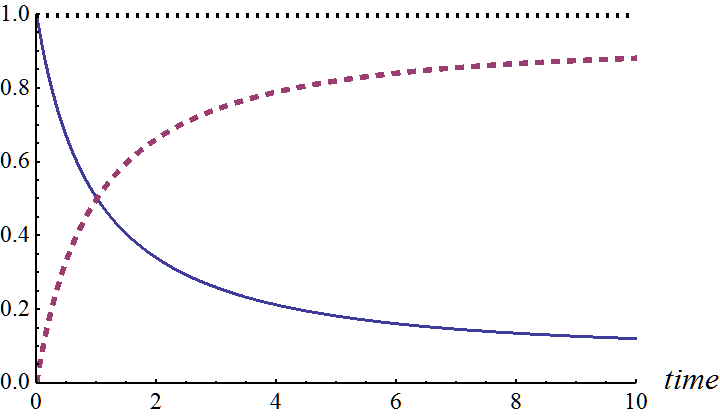}
\end{center}
\caption{Solid: laminar (linear) loss fraction of wave energy in the course of saturation, $f(t)=\tilde L(t)/(\tilde L(t)+\tilde P(t))$; 
dashed: ``wave age'', $1-f(t)$.}
\label{f:f_von_t-1}
\end{figure}

We note in passing that the steady state is not the only dynamically invariant 
state of TKE and waves. Also the state of \textit{exponential evolution} (Van Atta, 1999) 
in wind tunnels belongs to this class. This state, taken as a reference,
has also the property that perturbations relax back towards reference. 

In our present situation where we deal with relaxation times orders of magnitudes
apart, the use of the so-called Tikhonov principle seems to be helpful. 
It means to concentrate on 
processes with moderate relaxation times. (\ref{o-1}) is so fast that it can be taken as being 
always in structural equilibrium. (\ref{w-1}) is so slow that its time derivative is small compared 
with the source and sink terms at the right-hand side and can be neglected. We are thus left
with only one differential and two algebraic equations. However, in the case of
very stiff algebraic equations it is sometimes more useful to apply the method of non-stationary embedding.
Here it would mean to re-establish the character of  (\ref{o-1}) and (\ref{w-1}) as differential equations
and to seek the stationary solution via relaxation to the stationary state. 


\section{Special cases\label{balancetests}}

\noindent
Below we discuss some special cases of our general system(\ref{o-1} --\ref{w-1}):
the neutrally stratified case ($N=0$), the stratified ($N^2>0$) but geophysically shielded 
 ($\Pi=0$) case, and the stably stratified wind tunnel.

\noindent
\subsection{Neutral stratification, $N=0$\label{test21}}

\noindent
Homogeneous shear means constant shear along the horizontal and vertical axes.
The assumption $N=0$ means neutral stratification. Internal gravity waves 
of any kind are not supported by the fluid so that all terms
in (\ref{w-1}) vanish, together with this equation. Further, the terms $\Psi, \;\;\Pi$, 
$\tilde P$, $\tilde S$ are zero such that eventually (\ref{o-1}, \ref{k-1})
look as follows:
\begin{eqnarray}
\label{o-5}
	\frac{d \Omega}{d t}&=&  \frac{1}{\pi}\left(\frac{S^2}{2} - \Omega^2\right),\\
\label{k-5}
	\frac{d {\cal K}}{d t}&=& P - \varepsilon  .
\end{eqnarray}
These equations correspond to a clear mechanistic interpretation of turbulence as
dipole chaos in the sense of a two-fluid approach (excitons in form of quasi-rigid vortex tubes
made of inviscid fluid, and the materially identical inviscid but not excited fluid between the tubes)
derived by Baumert (2005 -- 2012). As a byproduct, this theory gives von Karman's number 
as $\kappa=1/\sqrt{2\,\pi} =0.399$. 


\subsection{Stable stratification, $N^2>0$, $\Pi=0$\label{test22}}

\noindent
Here we  mean stratified shear flows under idealized laboratory conditions where the role of 
tides and external geophysical influences are excluded, i.e.\ $\Pi=0$, which implies also $\tilde S=0$, to be discussed later. 
These conditions are called below ``geophysically shielded''. 

Homogeneity means here constant $N^2$ and $S^2$ in the horizontal plane and on the vertical axis, 
which is not easy to realize in a laboratory. But on a simplistic theoretical level we can get some insight 
when we consider only the stationary system (\ref{o-1}, \ref{k-1}, \ref{w-1}). 
The vorticity balance gives trivially $\Omega^2=S^2/2$ and is therefore 
omitted for brevity. It remains the following:
\begin{eqnarray}
\label{k-2}		0&=& P -\Psi -B - \varepsilon + \tilde P ,\\
\label{w-2}	0&=& \Psi-  \tilde L - \tilde P \,.
\end{eqnarray}
We neglect the linear molecular friction $\tilde L$, rewrite (\ref{w-2}) to get $\tilde P\approx \Psi$;
wie insert this  into (\ref{k-2}) and get the following:
\begin{equation}
\label{ocean-1}
	P\approx B + \varepsilon \,.
\end{equation}
Remarkably, $\Psi$ cancels out of (\ref{k-2}, \ref{w-2})
so that finally the oceanographer's standard balance formulation (\ref{ocean-1}) 
is obtained. We come back to this point later. 

In the past experiments with stratified wind tunnels gave deep insights into the nature of the
turbulence-wave interactions (Rohr et al., 1988; Van Atta, 1999; Baumert and Peters 2004, 2005). However, they do 
\textit{not} support relation (\ref{ocean-1}). This is the contradiction we mentioned in the Introduction
and  will be discussed next. 


\subsection{The stratified wind tunnel, $\Pi=0$, $R_g>0 $\label{test2}}
\subsubsection{The system.}
\noindent
This important case is an example of strong horizontal \textit{in}homogeneity and non-equilibrium conditions. 
In the entrance facility of the tunnel the forced flow passes a fine grid which leaves an initial high-frequency short-wave turbulence
and internal-wave signature in a locally homogeneous fluid body of limited size. We consider this fluid body as moving with the mean flow in a 
plug-flow sense. The small-scale properties then evolve during the trip within the fluid body 
along the homogeneously sheared and stratified tunnel until its end, where the body leaves the tunnel, including 
its small-scale properties. Typically an exponential evolution of TKE is observed along the longitudinal axis, 
either exponential growth or exponential decay (Van Atta, 1999). The waves were not recorded but it must be
hypothesized that they did reach saturation level.

This situation is artificial. In a natural hydrodynamic system the growth of turbulence is limited because it would somewhere
begin to reduce its own source (shear) by mixing the mean flow (see Fig.\ \ref{f:vorticity}) so that an equilibrium will sooner or later be reached, corresponding to
the large-scale energy input to the flow. But this feedback from the turbulence-wave system to the mean-flow system takes
time and the travel time through the tunnel is too short. 

Exactly this sort of decoupling of processes is what the tunnel experimentalists aim at. 
They wish in particular to cut off the feedback $\tilde P$ from shear-generated wave-energy
into TKE, as it contaminates the clear and simple picture. In other words, they want to see the naked 
interrelations in stratified shear turbulence as studied already by Richardson (1920), Howard (1961) and Miles (1961).  
Those three authors neglected shear-generated waves and their subsequent feedback as a result  of
spectral saturation. 

\subsubsection{Advection-dispersion-reaction (ADR) and the plug-flow concept\label{plug-flow}.}
At a first glance the sheared flow of a stratified wind tunnel seems to represent a major problem for detailed 
analyses. However, the Taylor-Aris theory of shear dispersion (Taylor, 1953; Aris, 1956; Baumert, 1973; 
Fischer et al., 1979) allows to compute an effective longitudinal dispersion coefficient, 
$D_L$, and to cast the transport equations corresponding to
(\ref{o-1}, \ref{k-1}, \ref{w-1}) in the following general form of an advection-dispersion-reaction equation:
\begin{equation}\label{generaldispersion}
\frac{\partial Y}{\partial t}+\frac{\partial }{\partial x}\left(\langle U\rangle \,Y-D_L\,\frac{\partial Y}{\partial x}\right) 	=  - Y/\hat \tau. 
\end{equation}
Here $Y$ is a placeholder for the variables $\Omega$, $\cal K$, and $\cal E$, and
$\hat \tau$ is an effective  time constant of a hypothetically decaying (or growing, when $\hat \tau<0$)  variable $Y$. 

As long as the Peclet number of the problem, $Pe=\left| \langle U\rangle^2 \hat \tau/(4\,D_L)\right| \gg 1$, it
can be shown (e.g.\ Baumert, 1973) that the stationary form of (\ref{generaldispersion}, with
${\partial Y}/{\partial t}=0$) can be simplified into a so-called plug-flow description: 
\begin{equation}\label{steadydispersion}
 \langle U\rangle \,\frac{d Y}{dx} \approx - Y/\hat \tau. 
\end{equation}
In a wind tunnel this concept is a useful approximation because the velocity $\langle U\rangle$ 
and thus the above similarity number are typically high enough. 
In a stationary plug-flow sense we thus have :
\begin{eqnarray}
\label{o-3}
	\langle U\rangle \frac{\partial\Omega}{\partial x}	
	&=&  \frac{1}{\pi}\left(\frac{S^2}{2} - \Omega^2\right),\\
\label{k-3}
	\langle U\rangle \frac{\partial\cal K}{\partial x}
	&=&P -\Psi -B - \varepsilon + \tilde P,   \\
\label{w-3}
	\langle U\rangle \frac{\partial\cal E}{\partial x}	
	&= &\Psi - \tilde L - \tilde P.
\end{eqnarray}
Now we introduce the \textit{dimensionless} travel-time coordinate along the wind-tunnel axis,
\begin{equation}\label{tprime}
	\hat t=x\cdot S/\langle U\rangle.
\end{equation}
We further introduce dimensionless variables via
\begin{eqnarray}
\label{o-non-1}
	\hat \Omega &=&  \Omega\, \sqrt{2}/S\,,\\
\label{k-non-1}
	\hat {\cal K}&=&{\cal K}/{\cal K}_0\,,   \\
\label{w-non-1}
	\hat {\cal E}&= &{\cal E}/{\cal E}_0\,.
\end{eqnarray}
Here ${\cal K}_0$ and ${\cal E}_0$ are the initial conditions so that\\
$\hat {\cal K}_{t=0}=\hat {\cal E}_{t=0}=1$. Due to the initiation of the
flow by the grid the vorticity typically begins with high initial values, $\hat\Omega_{t=0}\gg 1$.


\subsubsection{General case.}
The above conventions allow to rewrite the transport equations (\ref{o-3}, \ref{k-3}, \ref{w-3}) 
with some algebra in dimensionless form:
\begin{eqnarray}
\label{o-non-3}
	\frac{d\hat\Omega}{d\hat t}	
	&=&  \frac{1}{2\,\pi}\left(1 - \hat \Omega^2\right),\\
\label{k-non-3}
	\frac{d\hat {\cal K}}{d\hat t}
	&=&\left(2-\frac{2 R_g}{\sigma_0}-\hat\Omega^2\right)\frac{\hat {\cal K}}{\pi\,\hat\Omega\,\sqrt{2}} + \frac{\tilde P}{S\cdot {\cal K}_0},   \\
\label{w-non-3}
	\frac{d\hat {\cal E}}{d\hat t}	
	&= &\frac{{\cal K}_0}{{\cal E}_0}\left( \frac{\sqrt{8}}{\pi\,\sigma_0} \,R_g^2 \;\frac{\hat {\cal K}}{\hat\Omega^3}- \frac{\tilde P}{S\cdot {\cal K}_0} \right).
\end{eqnarray}
Here we neglected the linear molecular friction term $\tilde L = c_1\,\cal E$
and replaced according to (\ref{BW}) $\Psi$ with $\nu\,N^2/\sigma_0-B$. For $B$ we used (\ref{BB}) and (\ref{mu}). 
\begin{figure}[h] 
\begin{center} 
\includegraphics[width=7cm,height=4.5cm,keepaspectratio]{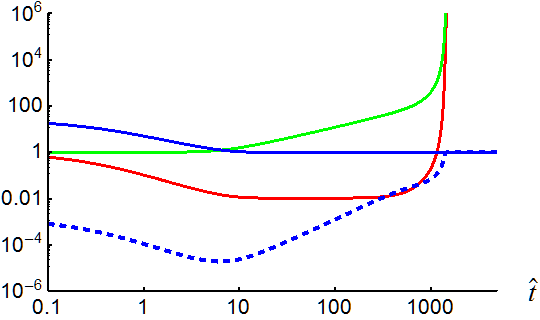}
\end{center}
\caption{Wind-tunnel turbulence-waves model in dimensionless variables. 
$\hat\Omega$ (solid blue) converges soon ($\hat t\approx 10$) to its \textit{structural-equilibrium} value, $\hat\Omega_{\infty}=1$. 
$\hat K$ (solid red) goes through a minimum and starts at $\hat t\approx 10$ a phase of exponential growth. 
$\hat {\cal E} $ (solid green) rests long time close to its initial condition $\hat {\cal E}_0=1$ and enters at $\hat t\approx 10$ 
into a phase of exponential growth. 
The ratio $\tilde P/\Psi$ (dashed blue, $0<\tilde P/\Psi <1$) remains initially very small but jumps then around $\hat t\approx 1500$ 
to its asymptotic value $(\tilde P/\Psi)_{\infty}=1.$ In this example $R_g=0.16$, $\hat \alpha=10^{-6}$, 
and $\beta=1$. 
Notice the double-logarithmic character of the presentation.}
\label{f:trans1}
\end{figure}
We now abbreviate $\alpha={\cal E}_0/S$, $\beta={\cal E}_0/{\cal K}_0$ and $\hat \alpha = \alpha\times c_2$ so that
\begin{eqnarray}
\label{o-non-4}
	\frac{d\hat\Omega}{d\hat t}	
	&=&  \frac{1}{2\,\pi}\left(1 - \hat \Omega^2\right),\\
\label{k-non-4}
	\frac{d\hat {\cal K}}{d\hat t}
	&=&\left(2-\frac{2R_g}{\sigma_0}-\hat\Omega^2\right)\frac{\hat {\cal K}}{\pi\,\hat\Omega\,\sqrt{2}} + \hat \alpha\,\beta\,\hat{\cal E}^2,   \\
\label{w-non-4}
	\frac{d\hat {\cal E}}{d\hat t}	
	&= &\frac{1}{\beta}\left( \frac{\sqrt{8}}{\pi\,\sigma_0} \,R_g^2 \;\frac{\hat {\cal K}}{\hat\Omega^3}-  \,\hat\alpha\,\beta\,\hat{\cal E}^2 \right).
\end{eqnarray}
According to Gregg (1989), in the ocean we have $c_2 = 7.4\times 10^{-5}$ s\, m$^{-2}$. $\alpha$ and $c_2$ are the only parameters or variables 
in (\ref{o-non-4} -- \ref{w-non-4}) which are \textit{not} dimensionless. But the two appear only in form of the product 
$\hat \alpha=\alpha\times c_2$, which is again dimensionless and was already used in (\ref{k-non-4}) and (\ref{w-non-4}).
To make an order of magnitude estimate of $\hat \alpha$ we take for $S$ the Garret-Munk value, $S_{GM} = 0.0036$ s$^{-1}$,
and $\hat {\cal E}\approx 10^{-4}$ m$^2$s$^{-2}$. This gives a characteristic guess of $ \hat \alpha \approx {\cal O}(10^{-6}$). 
This value was used to compute the data in Fig.\ \ref{f:trans1}.

The system (\ref{o-non-4} -- \ref{w-non-4}) can be solved numerically by standard methods if its high stiffness is adequately taken into 
account\footnote{Numerical overflows may occur. If automatic stiffness techniques are applied the solution may begin to  switch periodically
between different methods. It is always helpful to reduce the maximum time step as far as possible.}. 
The first phase of the evolution based on  (\ref{o-non-4} -- \ref{w-non-4})  is shown in Fig.\ \ref{f:trans1} where
we may identify three regimes with two separating breakpoints. 
The first regime is the initialization or spin-up regime. The first breakpoint
labels its end at $\hat t\approx 10$ and is associated with the transition into structural equilibrium. 
The second regime may be called the typical wind-tunnel regime (an artificially decoupled or naked turbulence-wave system in
exponential growth, $\tilde P\approx 0$). Its end is labeled by the second breakpoint at $\hat t\approx 1500$. 
The third regime can be called a hyper-equilibrium regime 
characterized by $\tilde P=\Psi$ and exponential growth, too. Both exponential growth regimes exhibit the property 
\begin{eqnarray}
	\label{hyper1}
	\frac{1}{\hat K}\frac{d\hat K}{d\hat t}&=& \mbox{constant}\,,\\
	\label{hyper2}
	0 &=&\frac{d}{d\hat t}\left(	\frac{1}{\hat K}\frac{d\hat K}{d\hat t}\right)\,.
\end{eqnarray}
We use for regime 3 the name \textit{hyper equilibrium} because it has the condition $\tilde P=\Psi$ 
in common with the natural equilibrium (oceanographer's regime) but differs with respect to
the dynamic state: while in the natural case all \textit{first} time derivatives vanish, in the hyper case 
they are constants $>0$, but the \textit{second} derivatives vanish.

The simulations show that in an ideal stratified wind tunnel of infinite length the hyper-equilibrium regime can in principle be reached, e.g.\ 
for $\hat t \gg 1500$. But such a length can hardly be achieved in practice. Thus the most interesting and scientifically unique part in a 
stratified wind tunnel (or a salt-stratified channel flow) is the section between the lower and the upper breakpoint.

In a somewhat sloppy form we may summarize: the total travel time through the wind tunnel is too short for the 
shear-generated waves to develop a saturated spectrum which would almost equate the breaking term $\tilde P$ with the
generation term $\Psi$. 


\section{Stably stratified natural shear flows: $R_g>0$, $\Pi>0, \tilde S^2>0$\label{Pig0}}

\noindent
We now come back to the system (\ref{o-1}, \ref{k-1}, \ref{w-1}) and study its the TKE balance where
the waves and the vorticity are assumed to stay in a steady state:
\begin{eqnarray}
\label{o-11}
	0 &=&  S^2/2- \Omega^2+ \tilde S^2,\\
\label{w-11}
	0&=&\Pi +\Psi-\tilde L - \tilde P\,,\\
\label{k-11}
	\frac{d\cal K}{dt}&=& P+ \tilde P - \Psi -B - \varepsilon \,.
\end{eqnarray}


\subsection{TKE balance and mixing efficiencies \label{tkebal}}

\noindent
The definition (\ref{f1}) allows now to rewrite the TKE balance (\ref{k-11}) as follows,
\begin{equation}	\label{tke-f1}
			\frac{d\cal K}{dt}=   (1-f)\Pi + P - B - \varepsilon - f\,\Psi\,.
\end{equation}
Shear and occasionally overturning waves are \textit{qualitatively} different generation mechanisms
with differing spectral signatures, differing mixing efficiencies and differing buoyancy fluxes. Both
mechanisms are associated with buoyancy fluxes which add up (parallel circuitry??) to the total buoyancy flux $B$:
\begin{eqnarray}\label{splitB}
	B&=& B'+\tilde B\,.
\end{eqnarray}
The so-called mixing efficiency, $\Gamma$, is generally defined as $\Gamma=B/\varepsilon$.
Due to the additive nature of $B=B'+\tilde B$ also $\Gamma$ is additive:
\begin{equation}	\label{def_10}
	\Gamma =\frac{B}{\varepsilon}\;=\frac{B'+\tilde B}{\varepsilon}\;=\;\frac{B'}{\varepsilon}+\frac{\tilde B}{\varepsilon}=	\Gamma'+\tilde \Gamma\,.
\end{equation}
With these preliminaries we may rewrite (\ref{tke-f1}) as follows:
\begin{equation}	\label{tke-f3}
			\frac{d\cal K}{dt}=   (1-f)\,\Pi + P - B'- \tilde B - \varepsilon - f\,\Psi\,.
\end{equation}


\subsubsection{Mixing efficiency of purely shear-generated turbulence: 
$\Gamma'=\Gamma'(R_g); \;\tilde B=0,\; \tilde P=0,\; \Pi=0,\;\tilde S^2=0.$ \label{waveless1}}
\begin{figure}[h] 
\begin{center} 
\includegraphics[width=7cm,height=4.5cm,keepaspectratio]{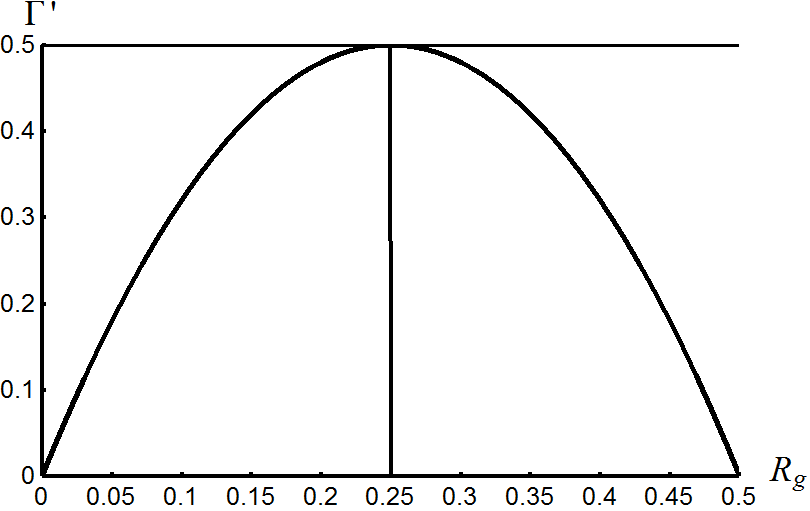}
\end{center}
\caption{The mixing efficiency $\Gamma'$ according to equation (\ref{def103}) for shear-generated turbulence.}
\label{f:gammas}\end{figure}

Using exclusively definitions like (\ref{eps}) and (\ref{prandtl1}) given in the previous sections and subsections, we begin here to specify $B'$ as follows:
\begin{equation}\label{def101}
	B'= \varepsilon\;\Gamma' =\mu\,N^2 = \frac{\nu}{\sigma}\,N^2=\frac{\nu}{\sigma_0} \left[ 1- \left( \frac{N}{\Omega} \right)^2\right]. 
\end{equation}
We solve (\ref{def101}) for $\Gamma'$ and get the following:
\begin{equation}	\label{def102}
	\Gamma'= \frac{\nu}{\sigma}\,\frac{N^2}{\varepsilon}=\frac{1}{\sigma} \left( \frac{N}{\Omega} \right)^2
	=\frac{1}{\sigma_0} \left( \frac{N}{\Omega} \right)^2 \left[ 1 - \left(  \frac{N}{\Omega} \right)^2 \right].
\end{equation}
In the pure shear-generated case, $\Pi = \tilde S=0$, we thus have $\Omega^2=S^2/2$, and with $R_g=N^2/S^2$ (\ref{def102}) rewrites as follows:
\begin{equation}	\label{def103}
	\Gamma'= \frac{2}{\sigma_0} R_g \left( 1 - 2\,R_g\right).
\end{equation}
This function is presented in Fig.\ \ref{f:gammas} and is well supported by observations as shown in an earlier study by Baumert and Peters (2004, 2005).

\subsubsection{Mixing efficiency of purely wave-generated turbulence: 
$\tilde \Gamma=\tilde \Gamma_0=0.2;\; P=\Psi=0,\; S=0.$\label{shearless1}}

The mixing situation is qualitatively different in the case without mean-flow shear where turbulence is exclusively generated by wave spectra 
fed eventually by large-scale long-wave external sources. According to the state of the art (Osborne, 1980; Oakey, 1982; Gregg et al., 1986; 
Peters et al. 1988), the mixing efficiency may still be taken to be a universal constant, $\tilde\Gamma =\tilde\Gamma_0=0.2$. Clearly, in purely 
shear-generated turbulence a gradient-Richardson number is not available so that $\tilde\Gamma$ cannot be a function of it.
Further, in this case this number is not even defined because the shear is zero.


\subsection{The remote-wave closure and oceanographer's balance: $P=\Psi=0,\; S=0,\; \Pi>0,\;\tilde S^2>0$\label{shearless2}}

\noindent
Here we will derive an approximate relation between the large-scale long-wave source of 
wave-generated turbulence (``remote waves''), $\Pi$, and the correspondingly induced component $\tilde S^2$ in the vorticity balance (\ref{o-11}).
We choose a flow without mean-flow shear, $S=0$ and $B'=0$, such that, according to (\ref{o-11}), $\Omega^2=\tilde S^2$. 
The TKE balance (\ref{tke-f3}) reads in this case 
\begin{equation}	\label{tke-f4}
	\frac{d\cal K}{dt}=   (1-f)\, \Pi-\tilde B -\varepsilon\,,
\end{equation}
and in the steady state:
\begin{equation}	\label{tke-f4}
	(1-  f)\, \Pi  =\tilde B + \varepsilon.
\end{equation}
The remaining rate $f\,\Pi$ describes the linear or molecular energy loss of the wave field.

We replace $\tilde B$ with $\tilde \Gamma\,\varepsilon$ and get
\begin{equation}	\label{tke-f5}
	 (1- f)\, \Pi = \tilde \gamma \,\varepsilon= \tilde \gamma \,\nu\,\Omega^2=\tilde \gamma \,\nu\,\tilde S^2
\end{equation}
so that the following is finally the closure of our problem:
\begin{equation}\label{ozzi22}
	\tilde S^2 =\frac{(1- f)\,\Pi}{\tilde \gamma \,\nu}\,.
\end{equation}


\section{Natural coexistence equilibria \label{pra1}}

\noindent
We consider now the general natural coexistence of waves and turbulence where $0\le f<1$, $\Pi>0$, $P>0$ etc.

\subsection{Vorticity \label{vortbal}}

\noindent
We insert the approximate closure relation (\ref{ozzi22}) in the vorticity balance (\ref{o-11}) and get with the 
abbreviation $\tilde\gamma = 1+\tilde\Gamma $
\begin{equation}\label{ozzi33}
	\Omega^2 = \frac{S^2}{2} +\frac{(1- f) \, \Pi}{\tilde\gamma\;\nu}\,.
\end{equation}
For convenience reasons we introduce $\eta$ and choose for (\ref{ozzi33}) the following presentation,
\begin{equation}\label{eta}
	\eta = \frac{\Omega^2}{S^2} = \frac{1}{2} +\frac{X}{\tilde \gamma } ,
\end{equation}
where for our convenience
\begin{equation}\label{ozzi332}
		 X= (1- f)\,\frac{\Pi}{\nu\,S^2}=(1- f)\,\frac{\Pi}{P}.
\end{equation}


\subsection{TKE\label{TKEBAL1}}

\noindent
Now we discuss the steady-state version of the general TKE balance (\ref{tke-f3}):
\begin{equation}	\label{tke-f33}
	(1- f)\, \Pi+P =B'+\tilde B +\varepsilon + f\, \Psi\,.
\end{equation}
According to our previous discussions, it may be written as follows: 
\begin{equation}	\label{tke-f34}
	(1- f)\, \Pi+\nu\,S^2 =\frac{\nu}{\sigma} \,N^2+\tilde \gamma\, \varepsilon + \frac{f}{\sigma_0}\left( \frac{N}{\Omega} \right)^4\varepsilon.
\end{equation}
We rewrite this equation identically as 
\begin{equation}	\label{tke-f35}
	(1- f)\, \Pi+\nu\,S^2 =\frac{\nu}{\sigma} \,N^2+\left[\tilde \gamma+\, \frac{f}{\sigma_0}\left( \frac{N}{\Omega} \right)^4\right]\nu\,\Omega^2,
\end{equation}
and divide both sides of (\ref{tke-f35}) by $\nu\,S^2$ to get with (\ref{ozzi332})
\begin{equation}	\label{tke-f36}
	1+X=\frac{R_g}{\sigma} +\left[\tilde \gamma+\, \frac{f}{\sigma_0}\left( \frac{N}{\Omega} \right)^4\right]\eta\;.
\end{equation}
We now expand $(N/\Omega)^4$ into $(N/S)^4/(\Omega/S)^4=R_g^2/\eta^2$, replace $\eta$ with
(\ref{eta})  and get finally
\begin{equation}	\label{tke-f37}
	\left( \frac{N}{\Omega}\right)^4 = \left( \frac{R_g}{ \eta  } \right)^2
	=\left(  \frac{2\,\tilde\gamma \,R_g}{\tilde \gamma +2\,X} \right)^2.
\end{equation}
Finally, with the helpf of (\ref{eta}), (\ref{tke-f37}) and with the abbreviation
 (\ref{tke-f36}) can be brought into the following final form:
\begin{eqnarray}	\label{tke-f38}
	1+X=\frac{R_g}{\sigma} +\left(\tilde \gamma +\, \frac{f}{\sigma_0}\;\frac{R_g^2}{\eta^2}\right)\,\eta\,,
\end{eqnarray}
so that 
\begin{eqnarray}	\label{tke-f381}
	1+X=\frac{R_g}{\sigma} + \tilde\gamma\,\eta +\, \frac{f}{\sigma_0}\;\frac{R_g^2}{\eta}\,,
\end{eqnarray}
where we remember that $\eta=1/2 + X/\tilde\gamma.$
We solve (\ref{tke-f381}) for $\sigma$:
\begin{eqnarray}	\label{tke-f382}
	\sigma_{(1)} =\frac{R_g}{1-\tilde\gamma/2 - 4\,\tilde\gamma\,f\, R_g^2 /(\tilde\gamma +2\,X)}\,.
\end{eqnarray}
This function should also satisfy the definition (\ref{prandtl1}) of the turbulent Prandtl number. This means that
\begin{equation}	\label{sig992}
		\sigma_{(2)}=\frac{\sigma_0}{1-(N/\Omega)^2}=\frac{\sigma_0\,\eta}{\eta-Rg}
\end{equation}
and with (\ref{eta}) we get
\begin{equation}	\label{sig993}
		\sigma=\frac{(\tilde\gamma+2\,X)\,\sigma_0}{\tilde\gamma+2\,X-2\,\tilde\gamma\,Rg}=\sigma_{(2)}(X, R_g)\,.
\end{equation}
Here $\sigma_0$ and $\tilde\gamma=1+\tilde\Gamma =1.2$ are a universal constants.

Now our unknown $X=(1-f)\,\Pi/P$ is easily determined by equating (\ref{tke-f381}) and (\ref{sig993}): 
\begin{equation}	\label{sig994}
		\sigma_{(1)}(f, X, R_g)=\sigma_{(2)}(X, R_g)\,.
\end{equation}
This gives the solution 
\begin{equation}	\label{sig995}
		X=X(f, R_g)\,.
\end{equation}
Now the knowledge of $X$ as function of $f$ and $R_g$ allows to 
present (\ref{sig993}) in the following form:
\begin{equation}	\label{sig9933}
		\sigma=\sigma_{(1)}(X_{(f, R_g)}, R_g)=\sigma_{(2)}(f, R_g)\,.
\end{equation}
The last expression is a family of curves giving us, in the way we are used to,
 for each value of $f \in (0, 1)$ \textit{one} curve $\sigma=\sigma(R_g)$ as function of the
gradient Richardson number.


\subsection{Validity limits\label{asymp2}}

\begin{figure}[thb] 
\begin{center} 
\includegraphics[width=8cm,height=5.cm,keepaspectratio]{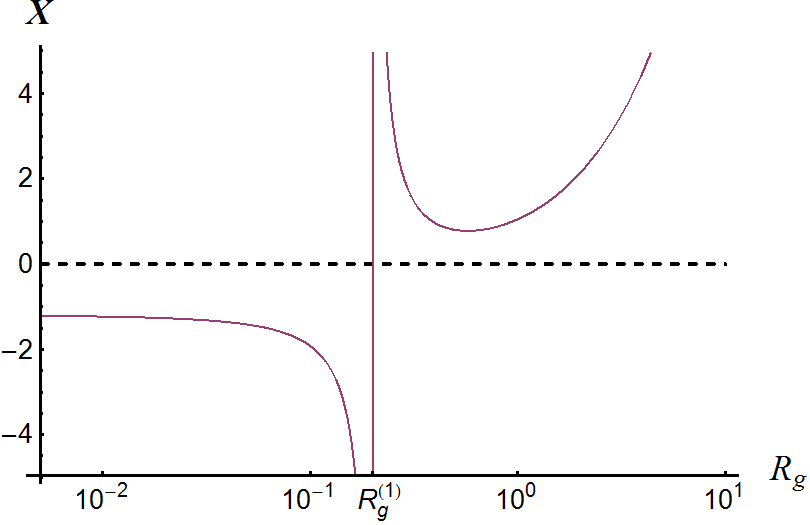}
\end{center}
\caption{The relative wave-energy  input to the TKE pool, $X=(1-f)\, \Pi/P$, as a function of the gradient Richardson number, $R_g$. 
Negative $X$ indicate withdrawal of energy from the wave-energy pool and are physically irrelevant. }
\label{f:asympt-1}\end{figure}

\noindent
For simplicity we analyze the system's behavior for $f=0$ and solve the following equation,
\begin{equation}	\label{sig9941}
		\sigma_{(1)}(f=0, X, R_g)=\sigma_{(2)}(X, R_g)\,,
\end{equation}
and find
\begin{equation}	\label{sig9942}
		\frac{2\,R_g}{2-\tilde \gamma} = \frac{(\gamma+2 X) \sigma_0}{\tilde\gamma+2 (X-\tilde\gamma Rg)}\, ,
\end{equation}
which is easily solved for $X$: 
\begin{equation}	\label{sig9943}
		X=\frac{1}{2}\; \frac{R_g-2\,\tilde\gamma\,R_g^2-\sigma_0\,(2-\tilde\gamma)\,\tilde\gamma}{\sigma_0\,(2-\tilde\gamma)/2-R_g}\,.
\end{equation}
Fig.\ \ref{f:asympt-1} illustrates the function $X=X(R_g)$ around the singular point, $R_g=  		R_g^{(1)}$:
\begin{equation}	\label{sig9944}
		R_g^{(1)} = (2-\tilde\gamma)\frac{\sigma_0}{2}=(1-\tilde\Gamma)\frac{\sigma_0}{2}=0.2\,.
\end{equation}
We see that 
\begin{equation}	\label{sig9945}
		X= \left\{ \begin{array}{ll}
				<0 & \mbox{if}\quad R_g <R_g^{(1)}, \\
				>0 & \mbox{if}\quad R_g >R_g^{(1)}. \\
					\end{array}\right. 
\end{equation}
We remember that according to (\ref{ozzi332}) negative $X \propto \Pi/P$ such that $X<0$ mean
sucking internal-wave energy out of the wave-energy pool. We accept therefore that  for 
$R_g <R_g^{(1)}$ a physically reasonable solution for an equilibrium coexistence 
of waves and turbulence does not exist.

The function $X$ in Fig.\ \ref{f:asympt-1} exhibits an obvious minimum. 
We take (\ref{sig9943}), differentiate and find  the zero of $dX/dR_g $ here:
\begin{equation}	\label{zero}
		R_g^{(2)}=(2-\tilde\gamma)\,\sigma_0= (1-\tilde\Gamma)\,\sigma_0=0.4. 
\end{equation}
Considering this point in Fig.\ \ref{f:asympt-1}, the solution on the left 
is physically unrealistic because a decreasing $R_g$  would lead to an increase in $X$. With decreasing
$R_g$ into a region left of the minimum in $X$ we approach a zone where the model is simply no longer correct. 
The reason is surely the vorticity balance which has been deduced from wind-tunnel experiments and then combined with
the new remote-wave closure (\ref{ozzi22}). Consequently its work is guaranteed only in two cases:
\begin{itemize}
  \item purely shear-generated turbulence with \textit{exclusion} of eigen-wave feedback, i.e.\ $\tilde P=0$ (wind tunnel);
  \item shear-generated turbulence with \textit{inclusion} of eigen-wave feedback, $\tilde P=(1-f) (\Pi+\Psi)>0$, 
	but dominant presence of remote waves, i.e.\ $X\propto \Pi/P \gg 1$  (ocean, atmosphere).
\end{itemize}  
Unfortunately, the important case with of eigen-wave feedback, $\tilde P=(1-f) \Psi>0$,  but without 
remote wave-energy source, $X\propto \Pi/P =0$, is not understood so far. It plays a role when
a flow setup is shielded from outer influences. This may play a role in technical systems like 
circulating cooling ponds where the waves stem not from geophysical sources and where they
evolve into a sufficiently saturated spectrum. This will be discussed in the next Section \ref{shield}.


\subsection{Asymptotic of $\sigma$  for
$R_g\rightarrow \infty. $  \label{asymp3}}

\noindent
We take the expression (\ref{sig9943}) for $X$ and look at very high $R_g$ where  in $R_g$ quadratic term
dominate constants and linear terms such that
\begin{equation}	\label{sig9946}
		\lim_{R_g\rightarrow\infty} X(R_g) = \frac{-2\,\tilde\gamma\,R_g^2}{-2\,R_g}={\tilde\gamma\,R_g}\,.
\end{equation}
The study of the Prandtl number function at high $R_g$ is still more easy. We take the left-hand side 
of (\ref{sig9942}), $\sigma_{(1)}=2\,R_g/(2-\tilde\gamma)$,
and use the definition $\tilde \gamma=1+\tilde\Gamma=1.2$ given already previously.
Obviously we have
\begin{equation}	\label{sig9947}
		\lim_{R_g\rightarrow\infty} \sigma(R_g) = \frac{2}{1-\tilde\Gamma}\,R_g=2.5\,R_g\;.
\end{equation}
For comparison with the numerical solution and with observational data we refer to Fig.\ \ref{f:naturalequilibrium-1}.

\section{Geophysically shielded systems\label{shield}}

\noindent
These systems have no external sources of internal-wave energy but they are
in coexistence equilibrium of turbulence and saturated waves with significant eigen-wave feedback, $f\ll 1 $.
Here $\Pi=0$ and the steady-state TKE balance reads in our notation as follows:
\begin{equation}	\label{s1}
		P=B'+\tilde B+\varepsilon\,+f\,\Psi\,=B'+\tilde\gamma \,\varepsilon\,+f\,\Psi\,.
\end{equation}
The vorticity balance is 
\begin{equation}	\label{s2}
		\eta=\frac{\varepsilon}{P}=\frac{\Omega^2}{S^2}= \frac{1}{2} +(1-f)\, \frac{Y}{\tilde\gamma}\,,
\end{equation}
where for young waves, $f\approx 1$, the effect of $Y$ is negligible and the situation is close to
the wind-tunnel case. Not so for older wave spectra. 

\begin{figure}[thb] 
\begin{center} 
\includegraphics[width=7cm,height=5.cm,keepaspectratio]{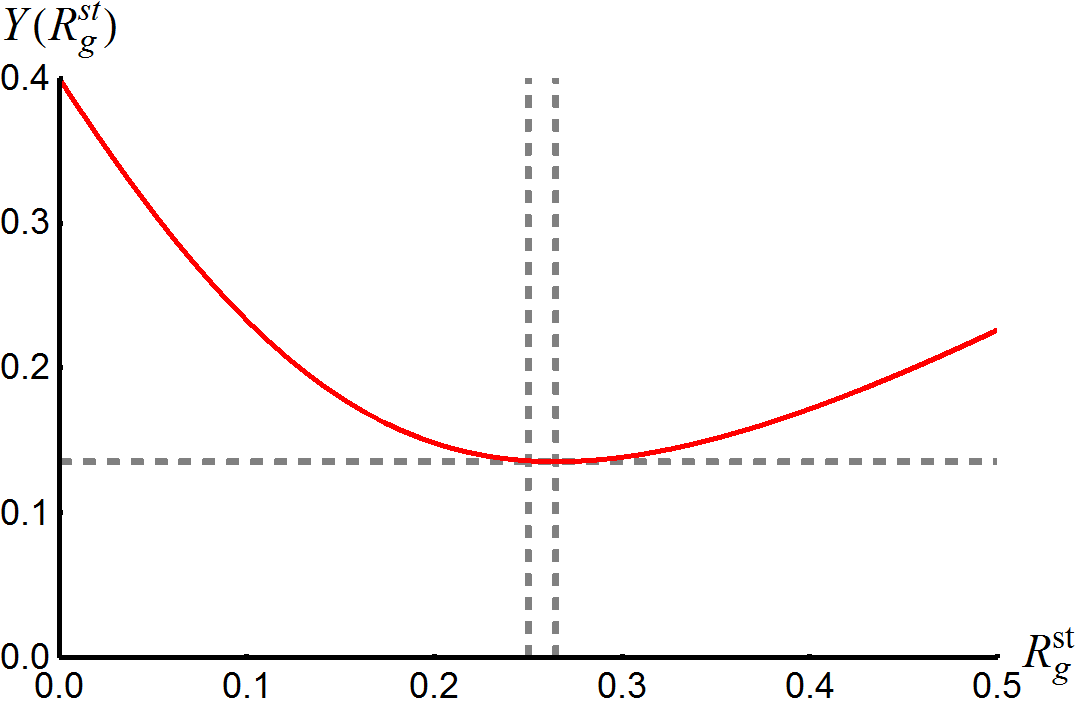}
\end{center}
\caption{The parameter $Y$ as a function of the steady-state gradient Richardson number $R_g^{st}$
and for $f=0$. 
The horizontal and the right vertical dashed gray lines cross at the minimum of the function $Y$. 
The left vertical dashed line labels $R_g^{st}=\onequarter$. 
Values $Y\approx 0.135 \dots 0.15$ 
would correspond to mathematically admissible values $R_g^{st}\approx 0.15 \dots 0.3$.}
\label{f:shieldY}\end{figure}

We combine (\ref{s1}) and (\ref{s2}) and get after some algebra the following 
steady-state condition,
\begin{equation}	\label{s3}
	2=4R_g\left(1-\frac{2\,\tilde\gamma\,R_g}{\tilde\gamma+2\,(1-f)\,Y}\right) +\tilde\gamma  +2\,(1-f)\,Y\,,
\end{equation}
 where $Y$ appears as a function of the steady-state
Richardson number $R_g^s$, and of the `wave age' $1-f$. The solution $Y=Y(R_g^s)$ of (\ref{s3}) is presented in
Fig.\ \ref{f:shieldY} for $f=0$. 

The above means that with $Y$ we have a tunable parameter which allows us to adjust
our model value for $R_g^s$ according to measurements or observations. Unfortunately these
are rare for shielded conditions described above so that we are inclined to choose
according to tradition the value $R_g^s=1/4$, which corresponds to $Y=0.136$.
Another choice would be the minimum value of $Y$ which is 0.135 and corresponds to
$R_g^s=0.265$. This somewhat arbotrary situation probably explains the large scatter
in the measurements of $\sigma$ and underlines umso mehr the necessity of 
dedicated experiments and observational programmes.

We note in closing this Section that the more general form of the steady-state vorticity balance is 
\begin{equation}	\label{s21}
		\eta=\frac{\varepsilon}{P}=\frac{\Omega^2}{S^2}= \frac{1}{2} + \frac{X+(1-f)\,Y}{\tilde\gamma}\,,
\end{equation}
where the external forces $X\propto \Pi/P$ appear together with the feedback via eigen waves, $Y$.
Our values of $Y$ are situated well below the validity limit of $X$. In the geophysically shielded 
case ($X=0$) and young waves ($f\approx 1$) the action of $Y$ is screened and we have $\eta=1/2$, 
which is the wind tunnel situation. In the same case with adult spectra ($f\ll 1$) we have 
the classical shielded case. 

\section{Application\label{app}}

\noindent
In the previous Sections of this report we have looked at special physical situations which we knew and
understood sufficiently well. Now we put these pieces together and write down the full equation system for applications
also to unknown situations. We choose here the philosophy of non-stationary embedding.
At a first glance the associated evolution equations (\ref{s3} -- \ref{s55}) look voluminous 
compared with the lean system of their purely algebraic steady-state counterparts, 
but non-stationary embedding avoids stiffness problems right on the most fundamental level and 
is thus substantially more robust in the computational practice. 


\subsection{Generalized equations}

\noindent
For an effective notation we define the following differential operator,
\begin{eqnarray}	\label{s3}
	{\cal D} &=& \left( \frac{\partial }{\partial t} - \frac{\partial}{\partial z} \,\nu \,\frac{\partial}{\partial z}\right),\\
	\tilde {\cal D} &=& \left( \frac{\partial }{\partial t} - \frac{\partial}{\partial z} \,\tilde\nu \,\frac{\partial}{\partial z}\right),
\end{eqnarray}
so that the general set of balances can be written for a stratified water column as follows:
\begin{eqnarray}	\label{o-33}
		{\cal D} \,\Omega&=&\frac{1}{\pi} \left[ \frac{S^2}{2}+\frac{1-f}{1+\tilde\Gamma}
	\left(\frac{\Pi}{P} + Y \right)S^2 -\Omega^2\right], \\
\tilde{\cal D}\,{\cal K} &=&\Pi+P-\nu\left[f\,\Psi \label{k-33}+
		 \frac{N^2}{\sigma}+(1+\tilde\Gamma)\Omega^2 \right]\\
\tilde	{\cal D}\,{\cal E} &=&\Pi+\Psi-c_1\,{\cal E} - c_2\,{\cal E}^2\;,\\
	f(t)&=&\frac{c_1\,{\cal E}}{c_1\,{\cal E} + c_2\,{\cal E}^2}\,,\\ 
	\sigma&=&\frac{\sigma_0}{1-N^2/\Omega^2}\;,\\
	P&=&\nu\,S^2\;,\\
	\Psi&=&\frac{P}{\sigma_0}\,\frac{S^2}{\Omega^2}\;R_g^2\,,\\
	\nu&=&{\cal K}/{\pi\,\Omega}\,,\\
	\mu&=&{\nu}/\sigma\,,\\
	R_g&=& N^2/S^2\,.\label{s55}
\end{eqnarray}
To apply this  theory in form of a numerical model 
it needs to be combined with a scheme which provides us with the mean-flow variables from which 
we may derive the shear $S$ and the Brunt-V\"ais\"al\"a frequency, $N$. Furthermore we need
initial conditions for all variables. But the hydrodynamic system is highly dissipative and `forgets' 
the initial conditions soon such that here ``reasonable guesses'' would suffice.


\subsection{The parameters}

\subsubsection{Overview.}
\noindent
$\tilde\Gamma=0.2$ and $\sigma_0=1/2$ are universal constants.
$f$ is a function of time. Its final equilibrium value, $f_{\infty}$, depends on the molecular-viscosity parameter $c_1$. 
In many cases it is sufficient to set $f_{\infty}=0$.

With the exception of $\tilde\nu$, the whole system contains only one tunable paramter, 
$Y\sim {\cal O}(10^{-1})$. The inner physical structure of $Y$ (and $\tilde\Gamma$) we 
have not yet fully understood. I.e.\ we are not able to derive its value from other than 
pragmatic arguments like `it works', because it  gives the right gradient Richardson number for the 
geophysically shielded steady-state.

About the spatial `diffusivity' of wave packets, $\tilde\nu$, we only know that it scales with the 
characteristic group velocity $\langle c_g\rangle$ of the wave packets, multiplied by a characteristic 
length scale which is possibly the characteristic wave lenght of a packet:
\begin{eqnarray}\label{calL}	
	\tilde\nu&\propto &\langle c_g\rangle \times \cal L\,. 
\end{eqnarray}
The last open problem to be discussed is the role of $\Pi$. According to Gregg (1989) it can be estimated from 
the r.m.s. 10-meter wave shear, $\langle S_{10}^2\rangle$, as function of time and of the region of the world ocean
under study. This will be done in the next Subsection.


\subsection{Wave-induced dissipation: the ocean case\label{waveind}}

\noindent
The above results apply to stratified oceanic and atmospheric flows as well.
The following estimator of the long-wave, non-local (``external'') energy source 
$\Pi$ is based on extensive studies in the world oceans and is thus not automtically 
applicable to atmospheric conditions. For the latter a comparable result is unknown.

Gregg (1989) presented a summary of comprehensive, extensive and \textit{direct} dissipation and 
stadardized shear observations ($S_{10}$) made in ocean waters around the globe, 
where the following conditions applied at least approximately:
\begin{itemize}
  \item There was almost no mean-flow shear, $S\approx 0$.
  \item The IGW field was almost perfectly saturated, $f\approx 0$. 
  \item The observations were done for conditions of quasi-steady state, $d \Omega/dt=d{\cal K}/dt=d{\cal E}/dt=0$.
\end{itemize}
Gregg established the following empirical relation between
the wave-induced dissipation rate, $\tilde\varepsilon$, the 10-meter high-pass filtered
vertical shear, $S_{10} = \Delta U/10\,\mbox{m}$, and the effective Brunt-V\"ais\"al\"a frequency, 
$\langle N^2\rangle^{1/2}$:
\begin{eqnarray}
	\label{g2}
	 \tilde\varepsilon&=& a_1 \, \frac{\langle N^2\rangle}{N_0^2} 
	\;\frac{\langle S_{10}^4\rangle}{S_{GM}^4}\; .
\end{eqnarray}
Very low frequencies have been removed from $S_{10}$  by filtering.
The average in $\langle S_{10}^4\rangle$ is taken over longer observation periods.

$S_{GM}$  used in (\ref{g2})  is the so-called Garrett-Munk shear:
\begin{eqnarray}
	\label{g3}
	S_{GM}^4 &=& a_2 \, \frac{\langle N^2\rangle^2}{N_0^4} ,
\end{eqnarray}
with the following empirical parameters:
\begin{eqnarray}
	\label{g4}
	a_1 &=& 7\times 10^{-10} \;\mbox{m$^2$\,s$^{-3}$}\,,\\
	a_2 &=& 1.66\times 10^{-10} \;\mbox{s$^{-4}$}\,,\\
	N_0 &=& 5.2\times 10^{-3} \;\mbox{s$^{-1}$}\,.
\end{eqnarray}
We insert (\ref{g3}) in (\ref{g2}) and get
\begin{eqnarray}
	\label{g5}
	\tilde\varepsilon &=& a_3 \, \frac{\langle S_{10}^4\rangle}{\langle N^2\rangle}\,,
\end{eqnarray}
where 
\begin{eqnarray}
	\label{g6}
	a_3 &=& \frac{a_1}{a_2}\,N_0^2 \;\;=\;\; 1.14\times 10^{-4} \;\mbox{m}^2\,\mbox{s}^{-1}\,.
\end{eqnarray}
We further take into account that  (see Gregg, 1989) 
\begin{eqnarray}
	\label{g7}
	\langle S_{10}^4\rangle&=& 2\, \langle S_{10}^2\rangle^2
\end{eqnarray}
such that (\ref{g5}) may be written as follows,
\begin{eqnarray}
	\label{g8}
	\tilde\varepsilon&=& a_4 \, \frac{\langle S_{10}^2\rangle^2}{\langle N^2\rangle}\;=\;a_4 \, \frac{\langle S_{10}^2\rangle}{\tilde R_g}\,,
\end{eqnarray}
with the wave-based gradient Richardson number, 
\begin{eqnarray} 
	\label{g9}
	\tilde R_g&=& \frac{\langle S_{10}^2\rangle}{\langle N^2\rangle}.
\end{eqnarray}
The latter needs to be unterschieden von the mean-field based gradient Richardson number 
$R_g=\langle N^2\rangle/\langle S\rangle ^2$. The parameter $a_4$ reads as follows:
\begin{eqnarray}
	\label{g11}
	a_4 &=& 2\, a_3\;\;=\;\; 2.3 \times 10^{-4} \;\mbox{m$^2$\,s$^{-1}$}\,.
\end{eqnarray}
With (\ref{g8}) we have a solid estimate of $\Pi$ as
\begin{eqnarray}
	\Pi=&=&(1+\tilde\Gamma)\,\varepsilon \;=\; \;a_4 \,(1+\tilde\Gamma)\, \frac{\langle S_{10}^2\rangle}{\tilde R_g}\,,
\end{eqnarray}
so that herewith the model system is completely closed and we can begin with simulations,
provided we have a certain knowledge about the wave-field parameters $\langle S_{10}^2\rangle$ and $\tilde R_g$
for our ocean or atmosphere region of interest.

\section{Comparison with observations\label{obs}}

\noindent
During the about 15-years walk towards the system (\ref{s3} -- \ref{s55}) 
partial solutions were step by step confronted with corresponding observational and experimental data. The totality of these
theory-observation comparisons cannot be repeated here. We refer for the wind-tunnel experiments
to Baumert and Peters (2004, 2005), for the Monin-Obukhov boundary layer to Baumert (2005a), for the
neutrally stratified case to Baumert (2005b, 2012). These specific cases are well described
by subsets of (\ref{s3} -- \ref{s55}), corresponding individually to the special physical situation studied. 

With respect to a theory-data comparison the present article  thus concentrates on the case $R_g\gg 1$
where without mentioning it $R_g$ means here always an equilibrium or steady-state value.
We look namely on the relation $\sigma=\sigma(R_g)$, analyzed and summarized by 
Zilitinkevich et al. (2008). These authors used more or less the same comprehensive data base for 
their discussions like Galperin et al. (2007), Canuto et al. (2008) and Kantha and Carniel (2009). 
In the core these are the CASES-99 (stable nocturnal BL, see Poulos et al., 2002) and the SHEBA experiments 
(arctic BL over ice; see Grachev et al., 2005, 2006). 

\begin{figure}[thb] 
\begin{center} 
\includegraphics[width=8cm,height=5.cm,keepaspectratio]{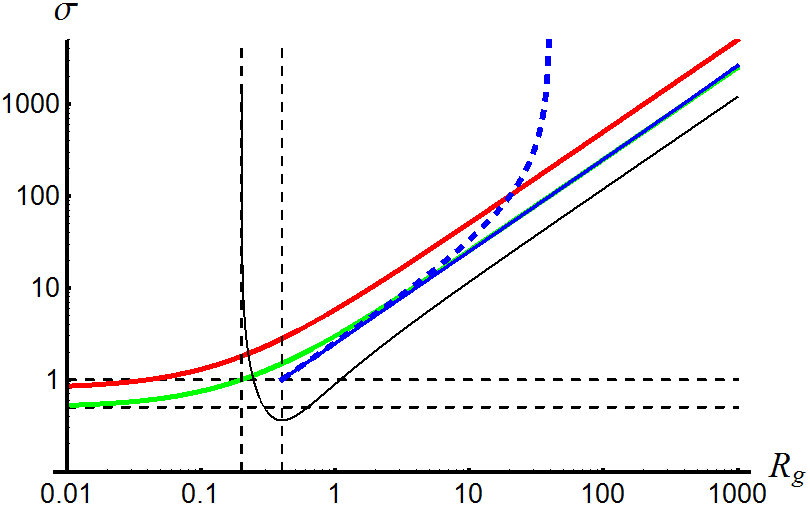}
\end{center}
\caption{Turbulent Prandtl number as function of the gradient Richardson number
for various coexistence equilibria of turbulence and internal waves.
The theory in dashed  blue  
contains the relative viscous energy loss of the wave field, $f$, taken here as $0.5\,\%$. 
Zilitinkevich et al. (2008) recommend as a phenomenological rule the relation 
$\sigma \approx 0.8+5\times\, R_g$ (full, red), which does not
contain the pole visible in the CASES-99 data and our $f=0.5\; \%$ case.
Only in the case $f=0$ our theory gives 
a straight line (full blue) which gives asymptotically $\sigma = (1 + 5\times R_g)/2$ (full green). 
The thin black dashed vertical lines indicate the positions 0.2 and 0.4. The thin black horizontal lines label 0.5 and 1.}
\label{f:naturalequilibrium-1}
\end{figure}

Zilitinkevich et al. (2008) conclude that the 
class-wise average of the data is best described  by $\sigma=0.8 + 5\, R_g$
shown in full red in Fig.\ \ref{f:naturalequilibrium-1}. In our view this line is somewhat above 
the CASES-99 data and our alternative  theoretically derived relation,  $\sigma=(1 + {5} \, R_g)/2$,
fits better. It is given in full green in Fig.\ \ref{f:naturalequilibrium-1} and located somewhat below the full-red curve. 
Of course, in view of the enormous scatter of the original data, the red and the green lines are well within the
huge overall scatter range. 

The two blue lines (full and dashed blue) in Fig.\ \ref{f:naturalequilibrium-1}  
represent our simulation results in good agreement with the observations of CASES-99.
At the same time they might illustrate \textit{why} natural data exhibit such a strong scatter. While the
dashed blue line is the case $f= 0.5\,\%$, the full blue line stands for $f=0$. I.e.\ very small
absolute variations in $f$ cause large variations in the solutions. Generally the huge scatter 
might be caused by the varying age (or degree of saturation) of the waves involved. These
waves are mostly of non-local origin so that in principle all points of the distant space are candidates
for their birth places and any age of arriving waves can be expected at our study site. 
On the one hand structural equilibrium needs to be achieved to avoid such a scatter.
This appears to be difficult under the action of waves which also modify the flow.
But still more important seems to be the different age of the incoming wave spectra.

According to our theory  Fig.\ \ref{f:naturalequilibrium-1}  shows also the 
two new critical gradient Richardson numbers for the wave-turbulence coexistence
in form of two vertical thin black dashed lines at $R_g^{(1)}=(1-\tilde\Gamma)/4=0.2$ 
and  $R_g^{(2)}=(1-\tilde\Gamma)/2=0.4$. The higher value labels the lower applicability limit of 
our theory. 

\section{Discussion}

\noindent
This report may be seen as an experiment to  draw a (dotted) theoretico-physical  bottom line under more than 50 years of 
ambitious observational programs and experimental laboratory research into 
\textit{stratified} small-scale geophysical and engineering turbulence and mixing processes. 
Looking backwards, today  these programs appear as a single planned 
initiative wherein mutually supplementing pieces fit perfectly together.
On the water side the efforts took place in the United States and the United Kingdom. 
On the atmospheric side also multi-national efforts need to be acknowledged. 
Our results may encourage those who believe that the ``turbulence problem'' 
(Hunt, 2011) is eventually solvable. 


\begin{acknowledgements} 
This study is a continuation of a long-term research engagement  in turbulence and mixing in oceans and inland waters,
initially supported by the European Union's projects PROVESS (MAS3-CT97-0159) and CARTUM (MAS3-CT98-0172), 
later by the U.S. National Science Foundation (OCE-9618287, OCE-9796016, OCE-98195056). The actual last phase was 
partially supported by the Department of the Navy Grant N62909-10-1-7050 issued by the Office of Naval Research Global. 
The United States Government has thus a royalty-free license throughout the world in all copyrightable material contained 
herein. Further partial support by the German BMBF, Ministry for Research and Education in the context oft he WISDOM-2 
project, is gratefully acknowledged.  The author further highly acknowledges the cooperation with Dr. Hartmut Peters from Earth and 
Space Research in Seattle, USA. His constant advice, corrective comments, interest and help made this study possible. 
\end{acknowledgements}

\nocite{zilitinkevichetal2008}
\nocite{galperinetal2007}
\nocite{canutoetal2008}
\nocite{kanthacarniel2009}
\nocite{rohrvanatta87}
\nocite{rohretal87}
\nocite{rohretal88a}
\nocite{rohretal88b}
\nocite{vanatta99}
\nocite{petersetal88}
\nocite{dasarolien00a}
\nocite{dasarolien00b}
\nocite{richardson1920}
\nocite{miles61}
\nocite{howard61}
\nocite{hazel72}
\nocite{thorpe1973}
\nocite{abarbaneletal84}
\nocite{baumertpeters04}
\nocite{baumertpeters05}
\nocite{baumert2012}
\nocite{gregg89}
\nocite{dickeymellor80}
\nocite{baumert05a}
\nocite{baumert05b}
\nocite{baumert73}
\nocite{fischeretal79}
\nocite{osborn80}
\nocite{oakey82}
\nocite{greggetal86}
\nocite{petersbaumert2007}
\nocite{munk2011}
\nocite{nakamuramahrt2005}
\nocite{grachevetal2005}
\nocite{grachevetal2006}
\nocite{grachevetal2007}
\nocite{baumertradach92}
\nocite{poulosetal2002}
\nocite{mahrt2006}
\nocite{itsweire84}
\nocite{itsweireetal86}
\nocite{hunt2011}
\nocite{aris56}
\nocite{taylor53}
\nocite{canuto2002}
\nocite{peters2008}
\nocite{woods2002}
\bibliography{vortexdipolchaos22,shewamix,p0,p1,p2,p34,p5,p6,p7a,p7b,p8,micro1,ocean1,tclos1,tclos2,ocean1,lake1,lake2,nature1}

\begin{thebibliography}{47}
\expandafter\ifx\csname natexlab\endcsname\relax\def\natexlab#1{#1}\fi
\expandafter\ifx\csname url\endcsname\relax
  \def\url#1{{\tt #1}}\fi
\expandafter\ifx\csname urlprefix\endcsname\relax\def\urlprefix{URL }\fi
\expandafter\ifx\csname doiprefix\endcsname\relax\def\doiprefix{doi:}\fi

\bibitem[{Abarbanel et~al.(1984)Abarbanel, Holm, Marsden, and
  Ratio}]{abarbaneletal84}
Abarbanel, H. D.~I., D.~D. Holm, J.~E. Marsden, and T.~Ratio, 1984: Richardson
  number criterion for the nonlinear stability of three-dimensional stratified
  flow. {\it Phys. Review Letter\/}, {\bf 52}, 2,352 -- 2,355.

\bibitem[{Aris(1956)}]{aris56}
Aris, R., 1956: On the dispersion of a solute in a fluid flowing through a
  tube. {\it Proc. Roy. Soc. A\/}, {\bf 235}, 67 --– 77.

\bibitem[{Baumert and Radach(1992)}]{baumertradach92}
Baumert, H. and G.~Radach, 1992: Hysteresis of turbulent kinetic energy in
  nonrotational tidal flows: {A} model study. {\it J. Geophys. Res.\/}, {\bf
  97}, 3669--3677.

\bibitem[{Baumert(1973)}]{baumert73}
Baumert, H.~Z., 1973: {\"U}ber systemtheoretische {M}odelle f{\"ur}
  {W}asserg{\"u}teprobleme in {F}lie{\ss}gew{\"a}ssern. {\it Acta
  Hydrophysica\/}, {\bf 18}, 5 -- 25.

\bibitem[{Baumert(2005a)}]{baumert05a}
Baumert, H.~Z.: 2005a, A novel two-equation turbulence closure for high
  {R}eynolds numbers. {Part B: S}patially non-uniform conditions. {\it Marine
  Turbulence: Theories, Observations and Models\/}, H.~Z. Baumert, J.~H.
  Simpson, and J.~S{\"u}ndermann, eds., Cambridge University Press, Chapter 4,
  31 -- 43.

\bibitem[{Baumert(2005b)}]{baumert05b}
--- 2005b, On some analogies between high-{R}eynolds number turbulence and a
  vortex gas for a simple flow configuration. {\it Marine Turbulence: Theories,
  Observations and Models\/}, H.~Z. Baumert, J.~H. Simpson, and
  J.~S{\"u}ndermann, eds., Cambridge University Press, Chapter 5, 44 -- 52.

\bibitem[{Baumert(2012)}]{baumert2012}
Baumert, H.~Z., 2012: Universal equations and constants of turbulent motion.
  {\it Physica Scripta, in press\/}.

\bibitem[{Baumert and Peters(2004)}]{baumertpeters04}
Baumert, H.~Z. and H.~Peters, 2004: Turbulence closure, steady state, and
  collapse into waves. {\it J. Phys. Oceanography\/}, {\bf 34}, 505 -- 512.

\bibitem[{Baumert and Peters(2005)}]{baumertpeters05}
Baumert, H.~Z. and H.~Peters: 2005, A novel two-equation turbulence closure for
  high reynolds numbers. part a: homogeneous, non-rotating stratified shear
  layers. {\it Marine Turbulence: Theories, Observations, and Models\/}, H.~Z.
  Baumert, J.~H. Simpson, and J.~S{\"u}ndermann, eds., Cambridge University
  Press, chapter~3, 14 -- 30.

\bibitem[{Canuto(2002)}]{canuto2002}
Canuto, V.~M., 2002: Critical {R}ichardson numbers and gravity waves. {\it
  Astronomy \& Astrophysics\/}, {\bf 384}, 1,119 -- 1,123.

\bibitem[{Canuto et~al.(2008)Canuto, Cheng, Howard, and Esau}]{canutoetal2008}
Canuto, V.~M., Y.~Cheng, A.~M. Howard, and I.~Esau, 2008: Stably stratified
  flows: {A} model with no {$Ri_{cr}$}. {\it J. Atmos. Sci.\/}, {\bf 65}, 2,437
  -- 2,447.

\bibitem[{D'Asaro and Lien(2000a)}]{dasarolien00a}
D'Asaro, E.~A. and R.-C. Lien, 2000a: {Lagrangian} measurements of waves and
  turbulence in stratified flows. {\it J. Phys. Oceanogr.\/}, {\bf 30}, 641 --
  655.

\bibitem[{{D'Asaro} and Lien(2000b)}]{dasarolien00b}
{D'Asaro}, E.~A. and R.~C. Lien, 2000b: The wave-turbulence transition for
  stratified flows. {\it J. Phys. Oceanogr.\/}, {\bf 30}, 1,669 -- 1,678.

\bibitem[{Dickey and Mellor(1980)}]{dickeymellor80}
Dickey, T.~D. and G.~L. Mellor, 1980: Decaying turbulence in neutral and
  stratified fluids. {\it J. Fluid Mech.\/}, {\bf 99}, 37 -- 48.

\bibitem[{Fischer et~al.(1979)Fischer, List, Koh, Imberger, and
  Brooks}]{fischeretal79}
Fischer, H.~B., E.~J. List, R.~C.~Y. Koh, J.~Imberger, and N.~H. Brooks, 1979:
  {\it Mixing in Inland and Coastal Waters\/}. Academic Press, New York,
  London, 483 pp. pp.

\bibitem[{Galperin et~al.(2007)Galperin, Sukoriansky, and
  Anderson}]{galperinetal2007}
Galperin, B., S.~Sukoriansky, and P.~S. Anderson, 2007: On the critical
  {R}ichardson number in stably stratified turbulence. {\it Atmos. Sci.
  Lett.\/}, {\bf 8}, 65 -- 69.

\bibitem[{Grachev et~al.(2006)Grachev, Andreas, Fairall, Guest, , and
  Persson}]{grachevetal2006}
Grachev, A.~A., E.~L. Andreas, C.~W. Fairall, P.~S. Guest, , and P.~O. Persson,
  2006: Sheba data flux-profile relationship in the stable atmospheric surface
  layer. {\it Boundary Layer Meteorol.\/}, {\bf 117}, 315 -- 333.

\bibitem[{Grachev et~al.(2007)Grachev, Andreas, Fairall, Guest, and
  Persson}]{grachevetal2007}
Grachev, A.~A., E.~L. Andreas, C.~W. Fairall, P.~S. Guest, and P.~O.~G.
  Persson, 2007: On the turbulent prandtl number in the stable atmospheric
  boundary layer. {\it Boundary-Layer Meteorol.\/}, {\bf 125}, 329 -- 341,
  doi:10.

\bibitem[{Grachev et~al.(2005)Grachev, Fairall, Persson, Andreas, and
  Guest}]{grachevetal2005}
Grachev, A.~A., C.~W. Fairall, P.~O. Persson, E.~L. Andreas, and P.~S. Guest,
  2005: Sheba boundary-layer scaling regimes. the sheba data. {\it Boundary
  Layer Meteorol.\/}, {\bf 116}, 201 -- 235.

\bibitem[{Gregg(1989)}]{gregg89}
Gregg, M.~C., 1989: Scaling of turbulent dissipation in the thermocline. {\it
  J. Geophys. Res.\/}, {\bf 94}, 9,686 -- 9,697.

\bibitem[{Gregg et~al.(1986)Gregg, d'Asaro, Shay, and Larson}]{greggetal86}
Gregg, M.~C., E.~A. d'Asaro, T.~J. Shay, and N.~Larson, 1986: Observations of
  persistent mixing and near-inertial waves. {\it J. Phys. Oceanogr.\/}, {\bf
  16}, 856--885.

\bibitem[{Hazel(1972)}]{hazel72}
Hazel, P., 1972: Numerical studies of the stability of inviscid stratified
  shear flows. {\it J. Fluid Mech.\/}, {\bf 51}, 39--61.

\bibitem[{Howard(1961)}]{howard61}
Howard, L., 1961: Note on a paper of {John Miles}. {\it J. Fluid Mech.\/}, {\bf
  10}, 509 -- 512.

\bibitem[{Hunt(2011)}]{hunt2011}
Hunt, J., 2011: The importance and fascination of turbulence. Public Evening
  Lecture, Old Library,\\ ERCOFTAC -- 13$^{th}$ Europ. Turbulence Conf.
  (ETC13), 12 -- 15 September 2011, Warsaw, Poland, {see
  http://etc13.fuw.edu.pl/speakers/public-evening-lecture}.

\bibitem[{Itsweire(1984)}]{itsweire84}
Itsweire, E.~C., 1984: Measurements of vertical overturns in a stably
  stratified turbulent flow. {\it Phys. Fluids\/}, {\bf 27}, 764--766.

\bibitem[{Itsweire et~al.(1986)Itsweire, Helland, and Atta}]{itsweireetal86}
Itsweire, E.~C., K.~N. Helland, and C.~W.~V. Atta, 1986: The evolution of
  grid-generated turbulence in a stably stratified fluid. {\it J. Fluid
  Mech.\/}, {\bf 162}, 299 -- 338.

\bibitem[{Kantha and Carniel(2009)}]{kanthacarniel2009}
Kantha, L. and S.~Carniel, 2009: A note on modeling mixing in stably stratified
  flows. {\it J. Phys. Oceanography\/}, {\bf 66}, 2,501 -- 2,505.

\bibitem[{Mahrt(2006)}]{mahrt2006}
Mahrt, L., 2006: The influence of small-scale nonstationarity on turbulent
  transport for stable stratification. {\it Boundary Layer Meteorol.\/}, 1 --
  24.

\bibitem[{Miles(1961)}]{miles61}
Miles, J.~W., 1961: On the stability of heterogeneous shear flows. {\it J.
  Fluid Mech.\/}, {\bf 10}, 496--508.

\bibitem[{Munk(1981)}]{munk2011}
Munk, W.: 1981, Internal waves and small-scale processes. {\it Evolution of
  Physical Oceanography\/}, B.~A. Warren and C.~Wunsch, eds., The MIT Press,
  264 -- 291.

\bibitem[{Nakamura and Mahrt(2005)}]{nakamuramahrt2005}
Nakamura, R. and L.~Mahrt, 2005: A study of intermittent turbulence with
  cases-99 tower measurments. {\it Boundary Layer Meteorol.\/}, {\bf 114}, 367
  -- 387.

\bibitem[{Oakey(1982)}]{oakey82}
Oakey, N.~S., 1982: Determination of the rate of dissipation of turbulent
  energy from simultaneous temperature and velocity shear microstructure
  measurements. {\it J. Phys. Oceanogr.\/}, {\bf 12}, 256--271.

\bibitem[{Osborn(1980)}]{osborn80}
Osborn, T.~R., 1980: Estimates of the local rate of vertocal diffusion from
  dissipaton measurements. {\it J. Phys. Oceanography\/}, 83 -- 89.

\bibitem[{Peters(2008)}]{peters2008}
Peters, H., 2008: pers. comm.

\bibitem[{Peters and Baumert(2007)}]{petersbaumert2007}
Peters, H. and H.~Z. Baumert, 2007: Validating a turbulence closure against
  estuarine microstructure measurements. {\it Ocean Modelling\/}, {\bf 19}, 183
  -- 203.

\bibitem[{Peters et~al.(1988)Peters, Gregg, and Toole}]{petersetal88}
Peters, H., M.~C. Gregg, and J.~M. Toole, 1988: On the parameterization of
  equatorial turbulence. {\it J. Geophys. Res.\/}, {\bf 93}, 1199--1218.

\bibitem[{Poulos et~al.(2002)Poulos, Blumen, Fritts, Lundquist, Sun, Burns,
  Nappo, Banta, Newsom, Cuxart, Terradellas, b.~Galsley, and
  Jensen}]{poulosetal2002}
Poulos, G.~S., W.~Blumen, D.~Fritts, J.~Lundquist, J.~Sun, S.~Burns, C.~Nappo,
  R.~Banta, R.~Newsom, J.~Cuxart, E.~Terradellas, b.~Galsley, and M.~Jensen,
  2002: Cases-99: A comprehensive investigation of the stable nocturnal
  boundary layer. {\it Bull. Amer. Meteorol. Soc.\/}, {\bf 81}, 757 -- 779.

\bibitem[{Richardson(1920)}]{richardson1920}
Richardson, L.~F., 1920: The supply of energy from and to atmospheric eddies.
  {\it Proc. R. Soc. London\/}, {\bf A 97}, 354 -- 373.

\bibitem[{Rohr et~al.(1987)Rohr, Helland, Itsweire, and {Van
  Atta}}]{rohretal87}
Rohr, J.~J., K.~N. Helland, E.~C. Itsweire, and C.~W. {Van Atta}: 1987,
  Turbulence in a stratified shear flow: {A} progress report. {\it Turbulent
  Shear Flows\/}, F.~Durst, ed., Springer, New York.

\bibitem[{Rohr et~al.(1988{\natexlab{a}})Rohr, Itsweire, Helland, and {Van
  Atta}}]{rohretal88b}
Rohr, J.~J., E.~C. Itsweire, K.~N. Helland, and C.~W. {Van Atta},
  1988{\natexlab{a}}: Growth and decay of turbulence in a stably stratified
  shear flow. {\it J. Fluid Mech.\/}, {\bf 195}, 77--111.

\bibitem[{Rohr et~al.(1988{\natexlab{b}})Rohr, Itsweire, Helland, and {Van
  Atta}}]{rohretal88a}
--- 1988{\natexlab{b}}: An investigation of growth of turbulence in a
  uniform-mean-shear flow. {\it J. Fluid Mech.\/}, {\bf 188}, 1--33.

\bibitem[{Rohr and {Van Atta}(1987)}]{rohrvanatta87}
Rohr, J.~J. and C.~W. {Van Atta}, 1987: Mixing efficiency in stably stratified
  growing turbulence. {\it J. Geophys. Res.\/}, {\bf 92}, 5481--5488.

\bibitem[{Taylor(1953)}]{taylor53}
Taylor, G.~I., 1953: Dispersion of soluble matter in solvent flowing slowly
  through a tube. {\it Proc. Roy. Soc. A\/}, {\bf 219}, 186 --– 203.

\bibitem[{Thorpe(1973)}]{thorpe1973}
Thorpe, S.~A., 1973: Experiments on instability and mixing in a stratified
  shear flow. {\it J. Fluid Mech.\/}, {\bf 61}, 731 -- 751.

\bibitem[{{Van Atta}(1999)}]{vanatta99}
{Van Atta}, C.~W., 1999: A generalized {O}sborn-{C}ox model for estimating
  fluxes in nonequilibrium stably stratified turbulent shear flows. {\it J.
  Marine Systems\/}, {\bf 21}, 103 -- 112.

\bibitem[{Woods(2002)}]{woods2002}
Woods, J.~A.: 2002, Laminar flow in the ocean {E}kman layer. {\it Meteorology
  at the Millenium\/}, R.~P. Pearce, ed., Academic Press, San Diego etc.,
  volume~83 of {\it Intl. Geophysics Series\/}, 220 -- 232.

\bibitem[{Zilitinkevich et~al.(2008)Zilitinkevich, Elperin, Kleeorin,
  Rogachevskii, Esau, Mauritsen, and Miles}]{zilitinkevichetal2008}
Zilitinkevich, S., T.~Elperin, N.~Kleeorin, I.~Rogachevskii, I.~Esau,
  T.~Mauritsen, and M.~W. Miles, 2008: Turbulence energetics in stably
  stratified geophysical flows: Strong and weak mixing regimes. {\it Quarterly
  J. R. Meteorological Soc.\/}, {\bf 134}, 793 -- 799.

\end{thebibliography}
\end{document}